\documentclass[10pt,a4paper]{article}
\usepackage[utf8]{inputenc}
\usepackage{cite}
\usepackage[english]{babel}
\usepackage{graphicx}
\usepackage{geometry}
\usepackage{amsbsy,todonotes,cancel, mathrsfs,soul,ulem}
\usepackage[colorlinks=true, pdfstartview=FitV, linkcolor=blue, citecolor=blue, urlcolor=blue]{hyperref}
\usepackage{cancel,ulem}
\usepackage{amsmath,framed}
\usepackage{amssymb}
\usepackage{amsthm}
\usepackage{comment}
\usepackage{hyperref}
\def \nn{\nonumber}
\definecolor {darkgreen}{rgb}{0,0.6,0}
\def\le{\left}
\def\eqref#1{(\ref{#1})}
\def\ri{\right}

\def \QED{\hfill $\blacksquare$\par \vskip 5pt}

\def\be{\begin{eqnarray}}
\def\ee{\end{eqnarray}}
\def \e{\rm e}
\def \im{\, {\rm Im}}

\def\wt{\widetilde}
\newtheorem{lemma}{Lemma}[section]
\newtheorem{theorem}[lemma]{Theorem}

\newtheorem{proposition}[lemma]{Proposition}
\newtheorem{definition}[lemma]{\mathscr{D}efinition}

\newtheorem{remark}[lemma]{Remark}
\newtheorem{problem}[lemma]{Problem}
\numberwithin{equation}{section}
\def \Tr{ {\rm Tr}}
\def\ov{\overline}
\def\id{\mathrm {Id}\,}
\def \pa{\partial}
\definecolor{shadecolor}{rgb}{0.95, 0.95, 0.46}
\def\bd{
\definecolor{shadecolor}{rgb}{0.99, 0.9, 0.86}
\begin{shaded}
\begin{definition}}
\def\ed{\end{definition}
\end{shaded}}
\def\bp{
\definecolor{shadecolor}{rgb}{0.9, 0.99, 0.86}
\begin{shaded}
\begin{proposition}}
\def\ep{\end{proposition}
\end{shaded}}

\def\bea#1\eea{\begin{align}#1\end{align}}

\def\beas{\begin{eqnarray*}}
\def\eeas{\end{eqnarray*}}
\def \pa{\partial}
\def\C{{\mathbb C}}

\def\R{{\mathbb R}}

\def\d{\mathrm d}

\def\1{{\bf 1}}

\title{Integrable operators,   $\ov\pa$-Problems,  KP and NLS hierarchy}

\date{}

\begin{document}
\maketitle
\begin{center}
M. Bertola$^{\dagger\star}$ \footnote{Marco.Bertola@concordia.ca} 
T. Grava $^{\ddagger\diamondsuit}$ \footnote{grava@sissa.it}
G. Orsatti $^{\ddagger}$ \footnote{gorsatti@sissa.it}
\\
\bigskip
\begin{minipage}{0.7\textwidth}
\begin{small}
\begin{enumerate}
\item [${\dagger}$] {\it  Department of Mathematics and
Statistics, Concordia University\\ 1455 de Maisonneuve W., Montr\'eal, Qu\'ebec,
Canada H3G 1M8} 
\item[${\ddagger}$] {\it SISSA, International School for Advanced Studies, via Bonomea 265, Trieste, Italy  and INFN sezione di Trieste }
\item[${\star}$] {\it Centre de recherches math\'ematiques,
Universit\'e de Montr\'eal\\ C.~P.~6128, succ. centre ville, Montr\'eal,
Qu\'ebec, Canada H3C 3J7}
\item[${\diamondsuit}$] {\it School of Mathematics, University of Bristol, Fry Building, Bristol,
BS8 1UG, UK}
\end{enumerate}
\end{small}
\end{minipage}
\end{center}

\begin{abstract}
We develop the theory of integrable operators $\mathcal{K}$ acting on a domain   
of the complex plane with smooth boundary
in analogy with  the theory of integrable operators acting on contours of the complex plane. 
 We show how the resolvent operator is obtained from the solution of a
 $\ov\pa$-problem in the complex plane. 
 When such a   $\ov\pa$-problem depends on auxiliary parameters we define its Malgrange one form   in analogy with the theory of  isomonodromic problems. 
 We  show that the Malgrange   one form is closed and  coincides with the exterior logarithmic differential of the  Hilbert-Carleman determinant of the operator $\mathcal{K}$.
 With suitable choices of the setup  we show that the Hilbert-Carleman determinant is a $\tau$-function of the  Kadomtsev-Petviashvili (KP) or  nonlinear Schr\"odinger   hierarchies.
\end{abstract}
\section{Introduction}
\label{sec:intro}
A key notion in the theory of solvable integrable systems is that of $\tau$--function, (see \cite{Harnad2021} for a comprehensive historical perspective) and  in many instances such $\tau$--functions coincide with  the Fredholm determinant of an appropriate  
integral  operator. 

 A  distinguished class of integral operators  are the so called  {\it integrable operators}:
the theory of such operators has its roots in the work of Jimbo et al. \cite{JMMS80}  that ultimately led to the construction by   Its, Izergin, Korepin and Slavnov \cite{IIKS} of a Riemann-Hilbert   problem to express the kernel of their resolvent operators.
Their original motivation  for studying these operators comes  from the theory of
 quantum integrable models. The theory of integrable operators later found  applications in many fields of mathematics such as   random matrices and integrable partial differential equations: for example the {\it gap probabilities} in determinantal random point processes (and more generally the generating function of occupation numbers) are expressible as  a Fredholm determinant \cite{Soshnikov}, \cite{Deift} and this is at the core of the celebrated Tracy-Widom distribution for fluctuations of the largest eigenvalue of a random matrix in a Gaussian Unitary Ensemble \cite{TracyWidom}.
An integrable operator is an  integral operator acting on  $L^2(\Sigma,|\d w|)\otimes \mathbb{C}^n$ of the form
\[
\mathcal{K}[v](z)=\int_{\Sigma}K(z,w)v(w)dw,\quad z\in\Sigma,
\]
where $\Sigma$ is some oriented contour in  the complex plane  and  the  kernel $K(z,w)\in Mat(n\times n, \mathbb{C})$ has a special form
\bea
  \label{eq:K_ker0}
  K(z,w) :&= \frac{f^T(z) \, g(w)}{z - w} , \quad f(z),g(z) \in Mat(r \times n, \mathbb{C}), 
\eea
where $f$ and $g$ are rectangular $r\times n$ matrices and for the time being we only assume that $f$ and $g$ are smooth
along the connected components of $\Sigma$.
The condition for $K$ to be nonsingular requires 
\[
   f^{T}(z) \, g(z) \equiv 0.
 \]
In the  most relevant applications,   the operators of the form~\eqref{eq:K_ker0}  are  trace class operators.  
 An  important observation in \cite{IIKS} is that the resolvent operator
  \begin{equation}
  \label{Res}
  \mathcal{R}=\mathcal{K}(\id-\mathcal{K})^{-1} = (\id-\mathcal K)^{-1} - \id,
  \end{equation}
  where $\id$ is the identity operator, is in the same class, namely
   \[
 \mathcal{R}[v](z)=\int_{\Sigma}R(z,w)v(w)dw,\quad z\in\Sigma,
 \]  
 where the resolvent kernel has also the form of an integrable operator:
 \bea
  \label{eq:R_ker0}
  R(z,w) :&= \frac{F^T(z) \, G(w)}{z - w} , \quad F(z),G(z) \in Mat(r \times n, \mathbb{C}).
\eea
Here  $F^T(z)=(\id-\mathcal{K})^{-1}f^T$ and $G=g(\id-\mathcal{K})^{-1}$,
where $(\id-\mathcal{K})^{-1}$ in the first relation is acting to the right while in the second relation its action is  to the left.
  Another crucial observation of \cite{IIKS} (see also the introduction of \cite{Its_Harnad}) is that the determination of $\mathcal{R}$ is equivalent to the solution of an associated Riemann-Hilbert   (RH) problem for a $r\times r$ matrix $\Gamma(z)$ 
    analytic in $\C\backslash \Sigma$ that satisfies the boundary value relation (sometimes referred to as ``jump relation'')
    \begin{equation}
    \label{RH0}
    \begin{split}
   & \Gamma_+(z)=\Gamma_-(z)M(z), \;\;z\in\Sigma, \quad M(z)=\1+2\pi i f(z)g^T(z)\\
    &\Gamma(z)\to\1,\quad \mbox{as $|z|\to\infty$}.
    \end{split}
    \end{equation}
    Here $\Gamma_\pm(z)$ denote the boundary values of the matrix $\Gamma(z)$ as $z$ approaches from the left and right of   the oriented contour $\Sigma$ and $\1$ is the identity matrix in $\mbox{Mat}(r\times r,\C)$. The matrices $F$ and $G$ that define the resolvent kernel \eqref{eq:R_ker0} are related to the solution  $\Gamma$ of the RH problem \eqref{RH0}  by the relation 
    \begin{equation}
    \label{FG}
    F(z)=\Gamma(z) f(z),\quad G(z)=(\Gamma(z)^T)^{-1}g(z).
     \end{equation}
   This  connection between the Fredholm determinant  and  the RH problem    has been exploited in several contexts where the kernel depends on large parameters and the study of the asymptotic behaviour  of the Fredholm determinant 
   is obtained via  the Deift–Zhou nonlinear steepest descent method of the corresponding  RH problem \cite{DZ}. 
    This analysis has been  successfully implemented  for $n=1$ and $r=2$  in a large  class of kernels     originating in random matrices, orthogonal polynomials, probability and partial differential equations see for example (see e.g. \cite{DZ,Harnad2021,BorodinDeift, DeiftItsKrasovsky, MR1060387}).

 The knowledge of the resolvent operator allows to write variational formulæ for the
  Fredholm determinant of the operator
$\id - \mathcal{K}$  as
\be
\label{eq:Jvar}
\delta  \log \det (\id - \mathcal{K}) =-\mbox{Tr} \bigl((\id - \mathcal{K})^{-1} \circ \delta \mathcal{K}\bigr)= -\mbox{Tr} \bigl((\id + \mathcal{R}) \circ \delta \mathcal{K}\bigr),
\ee
 where here and below  $\delta$ stands for  exterior total  differentiation  in the space of parameters.

There exist   Riemann Hilbert problems more general than \eqref{RH0} that   describe the inverse monodromy problem of a linear system of first order  ODEs in the complex plane: in those cases the deformation parameters can be introduced in such a way that the monodromy data do not depend on them. 
In this context,
the Kyoto school headed by Jimbo, Miwa and Ueno \cite{JMU} introduced the concept of   isomonodromic $\tau$-function starting from a differential  one form  $\omega$  (the Malgrange one form)  defined on the space of isomonodromic deformation parameters.
In several situations the isomonodromic $\tau$-function can be identified with the Fredholm determinant (possibly up to multiplication by explicit factors) of an operator of integrable form \cite{Bertola},\cite{CGL},\cite{GL}.
%

 An enlargement of the class of integrable operators~\eqref{eq:K_ker0} was studied  by Bertola and Cafasso   \cite{BC} who 
considered Hankel composition operators that have been reduced  to integrable operators in Fourier space.
 
  Recently Bothner \cite{BB} and A. Krajenbrink \cite{KlD}  enlarged the class of Hankel composition operators that    can be studied via Riemann-Hilbert problems.  Applications are obtained in \cite{BCT}, \cite{CCR}.
  
The common feature of all these works  is the appearance, in one way or another, of a Riemann--Hilbert problem, namely, a boundary value problem of a matrix with discontinuities across a contour (or union thereof) with boundary values related multiplicatively by a group-like element $M$ (the 	``jump matrix'') as  in \eqref{RH0}.
  
  The goal of the present manuscript is to enlarge the class of integrable operators by considering  operators   $\mathcal{K}$ acting on   $L^2(\mathscr{D},\d^2 w)\otimes \mathbb{C}^n$ where  $\mathscr{D}$  is   a bounded  domain   of the complex plane   with a  matrix  kernel $K(z,\ov z,w,\ov w)\in  Mat(n  \times n, \mathbb{C})$, namely 
  \bea
  \label{eq:K_ker}
&\mathcal{K}[v](z,\overline{z})=\iint_{\mathscr{D}}K(z,\ov z,w,\ov w)v(w)\frac{\d\ov w \wedge \d{w}}{2 i},\quad z,\overline{z}\in\mathscr{D}, \\
  &K(z,\ov z,w,\ov w) := \frac{f^T(z,\ov z) \, g(w,\ov w)}{z - w} , \nn \\
  \label{fTg}
   & f^{T}(z,\ov z) \, g(z,\ov z) \equiv 0 \equiv (\pa_{\bar{z}} f(z,\ov z))^{T} \, g(z,\ov z),\;\;\;\;f,g\in \mathcal{C}^\infty(\mathscr D, Mat(r  \times n, \mathbb{C})).
\eea
 Here and below,  instead of $\ov\pa$, we use the symbol $\pa_{\bar{z}}$ to specify the derivative with  respect to $\ov z$.   The dependence of $f$ and $g$ on $z$ and $\ov z$ is to remind the reader that $f$ and $g$ are  in  general smooth matrix functions on the complex plane. 
The kernel $ K(z,\ov z,w,\ov w)$ and the corresponding integral   operator $ \mathcal{K}$ is a Hilbert--Schmidt operator with a well--defined
and continuous value on the  diagonal in $\mathscr{D}\times \mathscr{D}$    and therefore  its  trace and   Fredholm   determinant are well defined (Proposition 3.11.2 and  Theorem 3.11.5\cite{Simon_2015}) as limits of finite rank operators, namely
\begin{equation}
\label{trace0}
\lim_{n\to\infty}\Tr(P_n \mathcal{K}P_n)=\iint_{\mathscr{D}}K(z,\ov z,z,\ov z,)\frac{\d\ov z \wedge \d{z}}{2 i},
\end{equation}
where $P_n$ is a finite rank projection  such that $\lim_{n\to\infty}P_n=\id$.
Furthermore the limit
\begin{equation}
\label{det0}
\lim_{n\to\infty} \det(\id-P_n\mathcal{K}P_n)
\end{equation}
exists.  If $\mathcal{K}$ is trace class, then $\Tr(\mathcal{K})$ and $\det(\id-\mathcal{K})$ coincide with the limits \eqref{trace0} and \eqref{det0} respectively.
 Our results are the following.
 \begin{itemize}
 \item    In Section~\ref{sec:K}  we show that  the resolvent of the integral operator $\id-\mathcal{K}$ is obtained through the solution  of a
   $\overline{\partial}$-Problem   (instead of a Riemann-Hilbert problem)
   for a matrix--valued  function $\Gamma$:
   \be
   \begin{split}
   \label{dbar0}
&\pa_{\bar{z}} \Gamma(z,\ov z) = \Gamma(z,\ov z) M(z,\ov z);\qquad \Gamma(z,\ov z) \underset{z \to \infty}{ \to } \1,\\
&M(z,\ov z)=\pi f(z,\ov z)g^T(z,\ov z)\chi_{\mathscr{D}}(z),
\end{split}
\ee
where $\chi_{\mathscr{D}}(z)$  is the characteristic function   of the domain $\mathscr{D}$.  Note that the matrix $M(z,\ov z)$ is nilpotent because of \eqref{fTg}.
  We show that the $\overline{\partial}$-Problem is solvable if and only if the operator $\id-\mathcal{K}$
 is invertible.  Furthermore we show, in analogy with  integrable operators defined on contours, 
  that  the kernel of the resolvent is
  \begin{align*}
 R(z,\ov z,w,\ov w)= \frac{F(z,\ov z)^T\, G(w,\ov w)}{z - w}, \quad F(z, \ov z) = \Gamma(z, \ov z) f(z, \ov z),\  \ G(z, \ov z) = \Gamma^{-1}(z,\ov z) g(z, \ov z)
\end{align*}
where $\Gamma$ solves the $\ov\pa$-problem \eqref{dbar0}.
 
 \item In Section~\ref{sec:fred}  we consider the regularized determinant (Hilbert-Carleman determinant, equation 7.8 \cite{Gohberg_2000})
 of the operator $\mathcal{K}$. This  is defined 
 as the Fredholm determinant of the trace class operator  $\mathcal{T}_{\mathcal{K}}:=\id-(\id- \mathcal{K})e^{\mathcal{K}}$, 
 namely
 \begin{equation}
 \label{det2}
 \text{{\rm det}}_2(\id-\mathcal{K}):=\det(\id-\mathcal{T}_{\mathcal{K}})=\det((\id- \mathcal{K})e^{\mathcal{K}}).
 \end{equation}
 Using the Jacobi variational formula~\eqref{eq:Jvar}  we show that 
 \bea
 \delta\log& \text{{\rm det}}_2(\id-\mathcal{K})=\omega ,
 \\
\label{defomega}&\omega:= -\iint_{\mathscr{D}}\Tr\left(\Gamma^{-1}(z)\pa_z\Gamma(z)\delta M(z)\right)\frac{\d\ov z \wedge \d{z}}{2\pi i},
 \eea
 The one form $\omega$ is shown to be closed. In analogy with the literature on Riemann--Hilbert problems on contours \cite{Bert10, Bertola} we call $\omega$   the Malgrange one form of the $\overline{\partial}$-Problem. The corresponding $\tau $ function of the $\overline{\partial}$-Problem is henceforth defined by 
  \[
  \delta \log \tau=\omega\,.
  \]
 Therefore we have that
  \begin{equation}
\label{tau0}
   \tau=\text{{\rm det}}_2(\id-\mathcal{K}).
   \end{equation}
   If the operator $\mathcal{K}$ is   of trace class, then  the  $\tau$-function can be expressed as a  Fredholm determinant by the relation
     \[
 \tau=\det(\id-\mathcal{K}){\rm e}^{\Tr(\mathcal{K})}.
  \]
We also show (Subsection \ref{secMal})  that the formula \eqref{defomega} defines a closed one--form under the less restrictive assumption that $M(z,\ov z)$  is traceless  but not nilpotent, and this allows us to define a $\tau$--function (up to multiplicative constants) of the $\ov\pa$-problem 
\begin{equation}
\label{dbar1}
\pa_{\bar{z}} \Gamma(z,\ov z) = \Gamma(z,\ov z) M(z,\ov z);\qquad \Gamma(z,\ov z) \underset{z \to \infty}{ \to } \1, 
\end{equation}
by the relation 
\be
\label{tau1}
\delta \log \tau =
 \omega.
\ee
Note that in this more general case the $\tau$-function of  the  $\ov\pa$-problem is well defined but it is not in general related to a
Hilbert-Carleman determinant  of some integral operator.
   \item Finally in Section~\ref{sect:tau}
 we use the results of the previous section  by considering  the    $\ov\pa$-problem \eqref{dbar1} 
where  $M$    is  a $2\times 2$ matrix of    the form
\[
  M(z,\bar{z},\pmb{t})= {\rm e}^{\frac{\xi(z,\pmb{t})}{2}\sigma_{3}}M_{0}(z,\bar{z}){\rm e}^{-\frac{\xi(z,\pmb{t})}{2}\sigma_{3}}\;\quad \sigma_3=\begin{bmatrix}1&0\\0&-1\end{bmatrix},
\]
where $\xi(z,\pmb{t})= \sum_{j=1}^{+\infty}z^{j}t_{j}$ and $M_{0}(z,\bar{z})$ is a traceless matrix compactly supported on $\mathscr{\mathscr{D}}$; we show that   the corresponding $\tau$-function \eqref{tau1}  of the $\ov\pa$-problem  \eqref{dbar1}  is a Kadomtsev-Petviashvili  (KP)  $\tau$-function, 
 namely it satisfies  Hirota bilinear relations for  the  KP hierarchy   (see e.g. \cite{Harnad2021}).
 

We then specialize  the matrix $M$  of the $\ov\pa$-problem in  \eqref{dbar1} to the nilpotent and traceless  form 
$$M(z,\ov z;x,t)=\pi  {\rm e}^{-i(zx +z^2 t)\sigma_3} f(z,\ov z)g^T(z,\ov z){\rm e}^{i(zx +z^2 t)\sigma_3},\quad x\in \R,\;\; t\geq 0  $$ with   \begin{equation}
  f(z,\ov z)= \frac{1}{\sqrt{\pi}}
  \begin{bmatrix}
    \sqrt{\beta(z, \ov z)} \chi_{\mathscr{D}}(z)\\
    -\sqrt{\beta^{*}(z, \ov z)}\chi_{\ov{\mathscr{D}}}(\ov z)
  \end{bmatrix}\;\;\;
  g(z, \ov z)= \frac{1}{\sqrt{\pi}}
  \begin{bmatrix}
    \sqrt{\beta^{*}(z, \ov z)}\chi_{\ov{\mathscr{D}}}(\ov z)\\
    \sqrt{\beta(z, \ov z)} \chi_{\mathscr{D}}(z)
  \end{bmatrix},
\end{equation}
where $\beta^{*}(z,\ov z)=\overline{\beta(\ov z,z)}$ is a smooth function  and $\chi_{\mathscr{D}}$, $\chi_{\ov{\mathscr{D}}}$ are respectively the characteristic functions of a simply connected domain $\mathscr{D}\subset \C^+$ and its conjugate $\ov{\mathscr{D}}$.  Here $\C^+$ is the upper half space.
We show that  the  $\tau$-function  of   the  $\ov\pa$-problem  \eqref{dbar1} is  the  $\tau$-function  for the focusing Nonlinear Schr\"odinger   (NLS) equation and 
 coincides with  Hilbert-Carleman determinant of the operator $\mathcal{K}$ with integrable kernel $K(z,\ov z,w,\ov w) = \frac{f^T(z,\ov z) \, g(w,\ov w)}{z - w} $, namely 
\begin{equation}
\label{solNLS}
\pa^{2}_{x} \log \tau(x,t) =\pa^{2}_{x}\log\text{{\rm det}}_2(\id-\mathcal{K})= |\psi(x,t)|^2, \nn
\end{equation}
where  the complex function  $\psi(x,t)$ solves the focusing Nonlinear Schr\"odinger equation (NLS)
 \be
  \label{eq:NLS}
  i \pa_t   \psi + \frac{1}{2} \pa^{2}_x \psi + |\psi|^{2}\psi=0. 
 \ee
While  we can write the solution  to the NLS equation in the form \eqref{solNLS},  the analytical properties of such  family of initial data and solutions (e.g. the long-time behaviour)  have  still to be  explored.  
It is shown in \cite{BGO2023} that  such family of initial data  naturally emerge in the limit of  an infinite number of solitons. 
 We also illustrate how, for a specific choice of the domain
$\mathscr{D}$ and of the function $\beta$, the $\ov\pa$-problem \eqref{dbar0} can be reduced to a standard Riemann-Hilbert problem.  
  \end{itemize}

\section{Integrable operators and $\ov\pa$-problems}
\label{sec:K}

Let $\mathscr{D} \subset \mathbb{C}$  be a  compact union of domains with smooth boundary  and denote by  $\mathcal{K}$ the integral operator acting on the space $L^{2}(\mathscr{D},\d^2 z)\otimes \C^n$ with a kernel $K(z,w)$ of the form
\begin{align}
  \label{eq:K_ker2}
  &K(z,\ov z,w,\ov w) := \frac{f^T(z,\ov z)  g(w,\ov w)}{z - w} , \quad f(z, \ov z),g(z,\ov z) \in {\rm Mat}(r \times n, \mathbb{C}),  \\
   \label{eq:K_ker3}
   & f^T(z,\ov z)\, g(z,\ov z) \equiv 0  \quad \mbox{and }\quad   (\pa_{\bar{z}} f(z,\ov z))^{T} \, g(z,\ov z) \equiv 0,\quad z,\ov z\in  \mathscr{D}.
\end{align}
Here the matrix-valued functions $f,g$ are assumed to be sufficiently smooth on $\mathscr D$ but no analyticity is required and for this  reason we indicate the dependence on both variables $z$ and $\ov z$. 
The vanishing requirements along the locus $z=w$ are sufficient to guarantee that  the kernel $K$ admits a well-defined value on the diagonal and it is continuous on $ \mathscr{D}\times  \mathscr{D}$ 
\be
\lim_{w\to z} K(z,\ov z,w,\ov w)=K(z,\ov z, z,\ov z) = \pa_{z} f^T(z,\ov z)\, g(z,\ov z).
\ee
We have emphasized that the kernel and the functions are not holomorphically dependent on the variables; that said, from now on we omit the explicit dependence on $\ov z$, trusting that the class of functions we are dealing with  will be clear by the context each time.
The operator $\mathcal{K}$  acts as follows on functions
\be
\label{Koper}
\mathcal{K}[\varphi](z):= \iint_{\mathscr{D}} K(z,w) \varphi (w) \frac {\d \ov w \wedge \d w}{2i}, \ \ \varphi\in L^{2}(\mathscr{D},\d^2 z)\otimes \C^n.
\ee
We introduce the following $\ov\pa$-problem for an $r\times r$ matrix-valued function $\Gamma(z,\ov z)$.  
\begin{problem}
\label{dbarproblem}Find a matrix-valued function $\Gamma(z,\ov z)\in GL_r(\mathbb C)$ such that 
\be
\label{dbarpb}
\pa_{\bar{z}} \Gamma(z) = \Gamma(z) M(z);\qquad \Gamma(z) \underset{z \to \infty}{ \to } \1
\ee
where $\1$ is the identity in $GL_r(\mathbb C)$ and 
\begin{equation}
  \label{eq:M}
M(z):= 
\left\{
\begin{array}{ll}
{\pi} f(z)g^T(z),& \mbox{for $z\in \mathscr{D}$},\\
&\\
0& \mbox{for $z\in\C\backslash \mathscr{D}$}.
\end{array}\right.
\end{equation}

\end{problem}
We first show that 
\begin{lemma}
  \label{lemma_1}
If a solution of the $\ov\pa$-problem \ref{dbarproblem} exists, it is unique. Furthermore $\det \Gamma(z) \equiv 1$. 
\end{lemma}
\noindent {\bf Proof.}
 If $\Gamma$ is a solution of the $\ov\pa$-problem \ref{dbarproblem}  then
%

\bea
\pa_{\bar{z}} \det \Gamma = \Tr \Big({\rm adj} (\Gamma)\pa_{\bar{z}} \Gamma\Big)=\Tr \Big({\rm adj} (\Gamma) \Gamma M\Big)
\eea
where ${\rm adj}(\Gamma)$ denotes the adjugate matrix (the transposed of the co-factor matrix). 
  Here $ \Tr$  denotes  the matrix trace. Now the product in the last formula yields  ${\rm adj}(\Gamma)\Gamma =( \det \Gamma)\1 $, 
so that 
\be
\pa_{\bar{z}} \det \Gamma =\det(\Gamma) \Tr ( M) = 0
\ee
where the last identity follows from the fact that $M$ is traceless because $\Tr ( M) =\Tr ( M^T)=0$.  Thus $\det \Gamma$ is an entire function  which tends to $\1$ at infinity, and hence  it  is identically equal to $1$ by Liouville's theorem. 

Now, if $\Gamma_1,\Gamma_2$ are two solutions, it follows easily that $R(z):= \Gamma_1 \Gamma_2^{-1}$ is an entire matrix-valued function which tends to the identity matrix $\1$ at infinity and hence, by Liouville's theorem $R(z)\equiv \1$, thus proving the uniqueness.
\QED
\begin{theorem}
The operator $\id-\mathcal K$   with $\mathcal K$ as in \eqref{Koper} is invertible in $L^2(\mathscr D, \d^2 z) \otimes \C^n$ if and only if the $\ov\pa$-problem \ref{dbarproblem} admits a solution.   The resolvent $\mathcal{R}$ of $\mathcal{K}$ has  kernel  given by:
  \begin{equation}
    \label{eq:Resol}
    R(z,w) := \frac{f^{T}(z)\Gamma^{T}(z)\le(\Gamma^{T}(w)\ri)^{-1}\!\!\! g(w)}{z - w},\quad (z,w)\in \mathscr D\times \mathscr D
  \end{equation}
  where $\Gamma(z)$ is a $r \times r$ matrix  that solves the $\ov\pa$-problem \ref{dbarproblem}.
\end{theorem}

\noindent {\bf Proof.}
Suppose that  the $\ov\pa$-problem~\ref{dbarproblem} is solved by $\Gamma(z)$; we now show that the operator  $(\id- \mathcal{K})$ is invertible.
 Let us define the  operator
  \begin{equation*}
    \mathcal{R} : L^{2}(\mathscr{D},\d^2 z)\otimes \C^n\to L^{2}(\mathscr{D},\d^2 z)\otimes \C^n
  \end{equation*}
  with  kernel  $R(z,w)$ given by \eqref{eq:Resol}.
 To verify that $\mathcal R$ is  the resolvent of the operator $\mathcal{K}$  we need to check  the following condition 
 \color{black}
  \begin{equation}
    \label{eq:resol-cond}
    \begin{split}
      (\id+ \mathcal{R}) \circ  &(\id- \mathcal{K})= \id \\
      &\Downarrow\\
      \mathcal{R} \circ \mathcal{K} &= \mathcal{R} - \mathcal{K}.
    \end{split}
  \end{equation}
To this end we compute the kernel of $\mathcal{R} \circ \mathcal{K}$  namely  
\bea
(R\circ K)(z,w)&:= \iint_{\mathscr{D}}R(z,\zeta)K(\zeta,w) \frac {\d \ov \zeta \wedge \d \zeta}{2i}
\nn\\
&= \iint_{\mathscr{D}}\frac{f^{T}(z)\Gamma^{T}(z)\overbrace{(\Gamma^{T}(\zeta))^{-1}g(\zeta)f^{T}(\zeta)}^{=-\frac 1\pi  \pa_{\bar{\zeta}} (\Gamma^T(\zeta))^{-1}}g(w)}{(z - \zeta)(\zeta - w)} \frac {\d \ov \zeta \wedge \d \zeta}{2i}
\nn \\
 \nn \\
&=
-\frac{f^{T}(z)\Gamma^{T}(z)}{z-w}\iint_{\mathscr{D}} \pa_{\bar{\zeta}} (\Gamma^T(\zeta))^{-1}\le(\frac{ 1}{z - \zeta}+\frac1{\zeta - w} \ri)\frac {\d \ov \zeta \wedge \d \zeta}{2i\pi}g(w).
  \label{21231}
\eea  
 If we consider the generalized Cauchy-Pompeiu formula for the matrix $(\Gamma^T(z))^{-1}$ we can express it in  integral form as 
%
%
 \begin{equation}
   \label{eq:Cauchy_G}
   (\Gamma^T(z))^{-1} =   \1 - \iint\limits_{\mathscr{D}}\frac{\pa_{\bar{\zeta}}(\Gamma^T(\zeta))^{-1}}{\zeta - z} \frac{\d \ov \zeta \wedge d \zeta}{2 \pi i},\quad z\in\C.
 \end{equation}
We substitute \eqref{eq:Cauchy_G} into \eqref{21231}:
\bea     
(R\circ K)(z,w)&=
-\frac{f^{T}(z)\Gamma^{T}(z)}{z-w}\Big(\left( (\Gamma^T(z))^{-1} - \1 \right)
-\left( (\Gamma^T(w))^{-1} - \1 \right)\Big)g(w)
\nn\\
&=-\frac{f^{T}(z)\Gamma^{T}(z)}{z-w} 
\Big((\Gamma^T(z))^{-1}-(\Gamma^T(w))^{-1} \Big)g(w)
\nn\\
&= \frac{f^{T}(z)\Gamma^T(z) (\Gamma^{T}(w))^{-1}g(w) }{z-w} -\frac{f^{T}(z)g(w) }{z-w} = R(z,w)-K(z,w).
\eea
This shows that indeed $\mathcal R$ satisfies the resolvent equation \eqref{eq:resol-cond} and hence the operator $\id - \mathcal{K}$ is invertible.
\\[10pt]
 Viceversa, let us now suppose that the operator $\id - \mathcal{K}$ is invertible and denote 
 $$
 \mathcal{R} =\Big(\id - \mathcal{K}\Big)^{-1}-\id.
 $$
 We now verify that $\mathcal R$ has kernel
 \begin{equation}
   \label{eq:resolv_2}
   R(z,w)= \frac{F(z)^T\, G(w)}{z - w}
 \end{equation}
 where the matrices $F(z)$ and $G(z)$ are defined as
 \begin{equation}
   \label{eq:def_2}
   \begin{split}
     F(z)&:= (\id - \mathcal{K}^{T})^{-1}[f](z)\\
     G(z)&:=  (\id - \mathcal{K})^{-1}[g](z),\\
   \end{split} 
 \end{equation}
 with the inverse applied to each entry (and the transposition $T$ acts on the matrix indices).
 Indeed we verify  the condition~\eqref{eq:resol-cond} with $R$ given by~\eqref{eq:resolv_2}:
 \begin{equation*}
   \begin{split}
(R\circ K)(z,w)=    & \iint_{\mathscr{D}} \frac{F^{T}(z)G(\zeta)f^{T}(\zeta)g(w)}{(z - \zeta)(\zeta - w)} 
\frac{\d \ov \zeta \wedge \d {\zeta}}{2 i}
 =\frac{1}{z - w} \left(\iint_{\mathscr{D}}  \frac{F^{T}(z)G(\zeta)f^{T}(\zeta)g(w)}{(z - \zeta)} 
 \frac{\d \ov \zeta \wedge \d {\zeta}}{2 i} + \right. \\
    & \left. + \iint_{\mathscr{D}}  \frac{F^{T}(z)G(\zeta)f^{T}(\zeta)g(w)}{(\zeta - w)} \frac{\d \ov \zeta \wedge \d {\zeta}}{2 i} \right)\\
    &= \frac{1}{z - w}\Big(\mathcal{R}[f^{T}](z)g(w) + F^{T}(z)\mathcal{K}[G](w)\Big).\\      
   \end{split}
 \end{equation*}
Adding and subtracting the kernels $K(z,w)$ and $R(z,w)$, we obtain
 \bea
(R\circ K)(z,w)=\frac{1}{z - w}\Big((\id +& \mathcal{R})[f^{T}](z)g(w) - F^{T}(z)(\id - \mathcal{K})[G](w) \Big) + \nn \\
    & + R(z,w) - K(z,w) .
    \label{212}
\eea
With the definitions~\eqref{eq:def_2} the contributions in the first line of \eqref{212}  cancel out and the condition~\eqref{eq:resol-cond} is satisfied.
To conclude the proof we need to verify that 
\begin{align}
F(z) = \Gamma(z) f(z),\  \ G(z) = \Gamma^{-1}(z) g(z)
\end{align}
where the matrix $\Gamma$  solves the $\ov\pa$-problem \ref{dbarproblem}. 
To this end, let us define the matrix $\wt {\Gamma}(z)$
 \begin{equation}
   \label{eq:tild}
   \wt{\Gamma}(z) := \1 - \iint_{\mathscr{D}} \frac{F(\zeta) \, g^T(\zeta)}{\zeta - z} \frac {\d \ov \zeta \wedge \d \zeta}{2i},\quad z\in\C.
 \end{equation}
From this definition  it follows that 
 \begin{equation}
   \begin{split}
     f^{T}(z)\wt{\Gamma}^{T}(z) &= f^{T}(z) - \iint_{\mathscr{D}}\frac{f^{T}(z)g(\zeta) F^{T}(\zeta)}{\zeta - z} \frac {\d \ov \zeta \wedge \d \zeta}{2i}\\
     &= f^{T}(z) + \mathcal{K}[F^{T}](z) \\
     &=f^{T}(z) + F^{T}(z) - (\id-\mathcal{K})[F^{T}](z)\\
     &=F^{T}(z)\\
   \end{split}
 \end{equation}
 which implies
 \begin{equation}
   \label{eq:res_2}
   F(z)= \wt{\Gamma}(z)f(z).
 \end{equation}
 We now substitute \eqref{eq:res_2} in the definition \eqref{eq:tild}:
 \begin{equation}
   \wt{\Gamma}(z) = \1 - \iint_{\mathscr{D}}\frac{\wt{\Gamma}(\zeta) f(\zeta)g^{T}(\zeta)}{\zeta -z} \frac {\d \ov \zeta \wedge \d \zeta}{2i}.
 \end{equation}
 Then, following the general Cauchy formula~\eqref{eq:Cauchy_G}, we find that the matrix $\tilde{\Gamma}(z)$ satisfies
 \begin{equation}
  \pa_{\bar{z}}\tilde{\Gamma}(z)= \pi\wt {\Gamma}(z)f(z)g^{T}(z).
\end{equation}
Finally, since the support $ \mathscr{D}$ of $M$ is compact, the equation  \eqref{eq:tild} implies that $\wt\Gamma$ is analytic outside of $ \mathscr{D}$ and tends to $\1$ as $|z|\to \infty$. Thus $\wt \Gamma$ solves the same $\ov\pa$-problem \ref{dbarproblem} and since the solution is unique, it must coincide with  $\Gamma$.
\QED

\section{The Fredholm determinant}
\label{sec:fred}

In Section \ref{sec:K} we have  linked the solution of the $\ov\pa$-problem \ref{dbarproblem} to the existence of the inverse  of $\id -\mathcal{K}$. 
From the conditions~\eqref{eq:K_ker2} we conclude  that $\mathcal K$ is a Hilbert-Schmidt operator with a well-defined and continuous diagonal in $\mathscr{D} \times \mathscr{D}$: according to \cite{Simon_2015} this is sufficient to  define the Fredholm determinant for the operator $ \id - \mathcal{K}$, as explained in the following remark.

\begin{remark}
  In general, for a Hilbert-Schmidt operator $\mathcal{A}$, the Fredholm determinant is not defined but we can still define a regularization  of it,  called the \textit{Hilbert-Carleman determinant}

\begin{equation}
  \label{eq:HC_det}
  \text{{\rm det}}_2(\id - \mathcal{A}):= \text{{\rm det}}\left((\id - \mathcal{A})\e^{\mathcal{A}}\right).
    \end{equation}
We observe that  $\text{{\rm det}}\left((\id - \mathcal{A})\e^{\mathcal{A}}\right)= \text{{\rm det}}\left(\id-\mathcal{T}_{\mathcal{A}}\right)$ with $\mathcal{T}_{\mathcal{A}}:=\id-(\id - \mathcal{A})\e^{\mathcal{A}}$ and the  operator $\mathcal{T}_{\mathcal{A}}$ is trace class  because it has the representation
\[
\mathcal{T}_{\mathcal{A}}=-\sum_{n=2}^\infty\frac{n-1}{n!}\mathcal{A}^n.
\]
If $\mathcal{A}$  is of trace class  we can rewrite  the Hilbert-Carleman determinant as
\begin{equation}
\text{{\rm det}}_2(\id - \mathcal{A})= \det(\id-\mathcal{A}){\rm e}^{\Tr(\mathcal{A})}.
\end{equation}
\color{black}
Moreover,  as for the Fredholm determinant, the Hilbert-Carleman determinant can be represented by a series
\begin{equation}
 \label{eq:HC_det1}
 \text{{\rm det}}_2(\id - \mathcal{A})=   1 + \sum_{n=2}^{\infty}\frac{(-1)^n}{n!}\Psi_{n}(\mathcal{A})
 \end{equation}
where $\Psi_n(\mathcal{A})$ is given by the \textit{Plemelj-Smithies formula}
\begin{equation*}
  \Psi_{n}(\mathcal{A})=\det
  \begin{bmatrix}
    0 & n-1 & 0 & \dots & 0 &0\\
    \Tr(\mathcal{A}^{2})&0& n-2 &\dots &0 &0\\
    \Tr(\mathcal{A}^{3}) & \Tr(\mathcal{A}^2) &0&\dots &0 &0\\
    . & . & . & . & . & .\\
    \Tr(\mathcal{A}^{n}) & \Tr(\mathcal{A}^{n-1}) & \Tr(\mathcal{A}^{n-2}) &\dots& \Tr(\mathcal{A}^2)&0\\ 
  \end{bmatrix}.
\end{equation*}
It is shown  that if $\mathcal{A}$ is Hilbert-Schmidt then~\eqref{eq:HC_det1} converges (\cite{Gohberg_2000}, Chapter 10, Theorem 3.1).

\end{remark}

Let us now  assume that $\mathcal{K}$ depends smoothly on parameters $\pmb{t}=(t_1,t_2,\dots,t_j,\dots) \text{ with } t_j \in \C \text{ , } \forall \, j\ge 1$:  we want to relate solutions of the $\ov\pa$-problem~\ref{dbarproblem} with the variational equations for the determinant.

\begin{proposition}
\label{prop3.2}
Let us suppose that the matrix $M(z,\ov z)$   in the $\ov\pa$-Problem~\ref{dbarproblem},  depends smoothly on  some parameters $\pmb{t}$, while remaining identically nilpotent.  Then the solution $\Gamma(z)$   of the $\ov\pa$-problem~\ref{dbarproblem}   is related to the logarithmic derivative of the Hilbert-Carleman  determinant   of $\id -\mathcal K$ as follows:
  \begin{equation}
   \label{eq:lemma_3}
  \delta \log\Big[ \text{{\rm det}}_2(\id - \mathcal{K})\Big]= -\iint_{\mathscr{D}}\Tr\le(\Gamma^{-1}(z)\pa_z\Gamma(z)\delta M(z)\ri)\frac{\d\ov z \wedge \d{z}}{2\pi i},
  \end{equation}
  where $\delta$  stands for the total differential in the space of parameters $\pmb{t}$.
\end{proposition}

\noindent {\bf Proof.}
Using the Jacobi variational formula~\eqref{eq:Jvar}, we can rewrite the LHS of~\eqref{eq:lemma_3} as

\begin{equation}
  \label{eq:J-var}
  \delta \log\Big[\text{{\rm det}}_2(\id - \mathcal{K})\Big]= \delta \log\Big[\text{{\rm det}}\left((\id - \mathcal{K})\e^{\mathcal{K}}\right)\Big]= 
    -\Tr\le( \mathcal{R}\circ \delta \mathcal{K}\ri),
\end{equation}
where $\mathcal{R}\circ \delta \mathcal{K}$  is a trace class operator, since it is the composition of two Hilbert-Schmidt operators.
Here $\Tr$ denotes the trace on the Hilbert space $ L^{2}(\mathscr{D},\d^2 z)\otimes \C^n$.
The composition of the two operators $\mathcal{R} \circ  \delta \mathcal{K}$  produces the kernel
\bea
 (R\circ\delta K)(z,w) &=\iint_{\mathscr{D}}\frac{f^{T}(z)\Gamma^{T}(z)(\Gamma^{T}(\zeta))^{-1}g(\zeta)\delta(f^{T}(x)g(w))}{(z-\zeta)(\zeta-w)} \frac {\d \ov \zeta \wedge \d \zeta}{2i}
    \nn
    \\
    &=\iint_{\mathscr{D}}\frac{f^{T}(z)\Gamma^{T}(z)(\Gamma^{T}(\zeta))^{-1}g(\zeta)f^{T}(\zeta)\delta g(w)}{(z-\zeta)(\zeta-w)} \frac {\d \ov \zeta \wedge \d \zeta}{2i} + \label{eq:I}
   \\
   &+\iint_{\mathscr{D}}\frac{f^{T}(z)\Gamma^{T}(z)(\Gamma^{T}(\zeta))^{-1}g(\zeta)\delta  f^{T}(\zeta)g(w)}{(z-\zeta)(\zeta-w)} \frac {\d \ov \zeta \wedge \d \zeta}{2i} \label{eq:II}
   \eea
   where we have omitted explicit notation of the dependence on $\pmb t$ of the functions $f,g, F, G, \Gamma$.

   We focus on the term in \eqref{eq:I}.
   Using the identity $\frac 1{(z-\zeta)(\zeta-w)} = \frac 1{z-w} \le(\frac 1 {z-\zeta}+\frac 1{\zeta-w}\ri)$, we obtain 
\bea
\eqref{eq:I}= & \frac{f^{T}(z)\Gamma^{T}(z)}{z - w} \left(\iint_{\mathscr{D}} (\Gamma^{T}(\zeta))^{-1}g(\zeta)f^{T}(\zeta)\le(\frac 1 {z-\zeta}+\frac 1{\zeta-w}\ri)\frac {\d \ov \zeta \wedge \d \zeta}{2i}\right)\delta   g(w) 
    \label{32}
\eea  
In order to compute the trace we need to compute the kernel \eqref{32} along the diagonal $z=w$ and hence we consider  $\lim_{w\to z} \eqref{32}$.
Observe  that $(\Gamma^{T}(\zeta))^{-1}g(\zeta)f^{T}(\zeta) = -\frac 1 \pi \pa_{\bar{\zeta}} (\Gamma^T(\zeta))^{-1}$, and hence we can apply  the formula \eqref{eq:Cauchy_G} to eliminate the integral and rewrite \eqref{32} as follows
\bea
\eqref{32}&=-\frac{f^{T}(z)\Gamma^{T}(z)\Big((\Gamma^{T}(z))^{-1}-\1 \Big)\delta   g(w)}{z - w} +\frac{f^{T}(z)\Gamma^{T}(z)\Big( (\Gamma^{T}(w))^{-1}-\1 \Big)\delta  g(w)}{z - w}\\
&=\frac{f^{T}(z)\Big(\Gamma^{T}(z)(\Gamma^{T}(w))^{-1} -\1 \Big)\delta   g(w)}{z-w} \label{39}.
\eea
 We can now easily compute the expansion of~\eqref{39}  along the diagonal $w\to z$ by Taylor's formula, keeping in mind that $\Gamma$ is not a holomorphic function inside $\mathscr D$: 
\bea
\eqref{39}&=f^{T}(z) \pa_{z}\Gamma^T(z) (\Gamma^{T}(z))^{-1}\delta   g(z) + \frac{\bar{z} - \bar{w}}{z - w}\overbrace{f^{T}(z)g(z)}^{\equiv \, 0}f^{T}(z)\delta  g(w) + \mathcal{O}(|z - w|) \\
\nn \\
&=f^{T}(z) \pa_{z}\Gamma^T(z) (\Gamma^{T}(z))^{-1}\delta   g(z) + \mathcal{O}(|z - w|). 
 \eea
 Using the above expression we conclude that the  trace in $L^{2}(\mathscr{D},\d^2 z)\otimes \C^n$ of~\eqref{eq:I} is 
\begin{equation}
  \label{eq:Tr_I}
\Tr(\eqref{eq:I})=  \iint_{\mathscr{D}}
\Tr
\big(f^{T}(z) \partial_z\Gamma(z)^{T}(\Gamma^{T}(z))^{-1}\delta  g(z)\big)
\frac {\d \ov z\wedge \d z}{2i}.
\end{equation} 
Using the cyclicity of the trace and its invariance under transposition of the arguments, we reorder the terms~\eqref{eq:Tr_I}  to the form 
\begin{equation}
  \label{eq:eq:Tr_I_fin}
\Tr(\eqref{eq:I})=\iint_{\mathscr{D}}\Tr\big(
\Gamma^{-1}(z)\partial_z\Gamma(z)f(z)\delta  g^{T}(z)
\big)
\frac {\d \ov z\wedge \d z}{2i}.
\end{equation}
We now consider the term \eqref{eq:II}.  Taking its trace yields:
\bea
  \label{eq:Tr_II}
    &\Tr(\eqref{eq:II})=\nn
    \\
   & =-\iint_{\mathscr{D}}\iint_{\mathscr{D}}\frac{\Tr\big(f^{T}(z)\Gamma^{T}(z)(\Gamma^{T}(\zeta))^{-1}g(\zeta)\delta  f^{T}(\zeta)g(z)\big)}{(z - \zeta)^{2}}\frac {\d \ov \zeta \wedge \d \zeta}{2i} \frac {\d \ov z \wedge \d z}{2i} 
      \nn \\
  & = -\iint_{\mathscr{D}}\iint_{\mathscr{D}}\frac{\Tr\big(g(z)f^{T}(z)\Gamma^{T}(z)(\Gamma^{T}(\zeta))^{-1}g(\zeta)\delta  f^{T}(\zeta)\big)}{(z - \zeta)^{2}}\frac {\d \ov \zeta \wedge \d \zeta}{2i} \frac {\d \ov z \wedge \d z}{2i} 
\eea
 We observe that the integrand is  in  $L^2_{loc}$ because the numerator vanishes to order $\mathcal O(|z-\zeta|)$ along the diagonal
\bea
\Tr\Big(g(z)f^{T}(z)\Gamma^{T}(z)(\Gamma^{T}(\zeta))^{-1}g(\zeta)\delta  f^{T}(\zeta)\Big)
=
\Tr\Big(g(\zeta)\overbrace{f^{T}(\zeta)g(\zeta)}^{=0}\delta  f^{T}(\zeta)\Big)
+\mathcal O(|z-\zeta|),
\eea
and hence the integrand is $\mathcal O(|z-\zeta|^{-1})$ which is locally integrable with respect to the area measure.
 We can now relate this integral to  $\pa_z\Gamma$ as follows.
Using the formula~\eqref{eq:Cauchy_G} and the $\ov\pa$-problem \ref{dbarproblem}  we can rewrite $\Gamma^{T}(\zeta)$ as
\begin{equation}
  \label{eq:G-1}
  \Gamma^{T}(\zeta)= \1 - 
  \iint_{\mathscr{D}}\frac{\pa_{\bar{z}}(\Gamma^{T}(z))}{z - \zeta} \frac{\d\ov z\wedge \d{z}}{2 \pi i} =  \1
  - \iint_{\mathscr{D}}\frac{M^{T}(z)\Gamma^{T}(z)}{z - \zeta} \frac{\d\ov z\wedge \d{z}}{2 \pi i}
\end{equation}
%
 Taking the holomorphic derivative  with respect to $\zeta$ we get
 \begin{equation*}
   \partial_\zeta\Gamma^{T}(\zeta)= -\iint_{\mathscr{D}}\frac{g(z)f^{T}(z)\Gamma^{T}(z)}{(z-\zeta)^{2}}\frac{\d \ov z\wedge \d {z}}{2 i}
 \end{equation*}
Plugging the result  into \eqref{eq:Tr_II} we obtain
 \begin{equation}
   \begin{split}
     \eqref{eq:Tr_II}=&-\iint_{\mathscr{D}}\Tr\le((\Gamma^{T}(\zeta))^{-1}g(\zeta)\delta   f(\zeta)\left(\iint_{\mathscr{D}}\frac{g(z)f^{T}(z)\Gamma^{T}(z)}{(z -\zeta)^{2}}\frac {\d \ov z \wedge \d z}{2i}\right)\ri) \frac {\d \ov \zeta \wedge \d \zeta}{2i}\\
     = &\iint_{\mathscr{D}}\Tr\big((\Gamma^{T}(\zeta))^{-1}g(\zeta)\delta  f^{T}(\zeta)\partial_\zeta(\Gamma^{T}(\zeta))\big)\frac {\d \ov \zeta \wedge \d \zeta}{2i} \\
     =&\iint_{\mathscr{D}}\Tr\big(g(\zeta)\delta  f^{T}(\zeta)\partial_\zeta\Gamma^{T}(\zeta)(\Gamma^{-1}(\zeta))^{T}\big)\frac {\d \ov \zeta \wedge \d \zeta}{2i}\\
     = &\iint_{\mathscr{D}}\Tr\big(\Gamma^{-1}(\zeta)\partial_\zeta\Gamma(\zeta)\delta  f(\zeta)g^{T}(\zeta)\big)\frac {\d \ov \zeta \wedge \d \zeta}{2i},
   \end{split}
 \end{equation}
 so that 
 \begin{equation}
 \label{part2}
  \Tr(\eqref{eq:II})=\iint_{\mathscr{D}}\Tr\big(\Gamma^{-1}(\zeta)\partial_\zeta \Gamma(\zeta)\delta  f(\zeta)g^{T}(\zeta)\big)\frac {\d \ov \zeta \wedge \d \zeta}{2i},
  \end{equation}
Combining \eqref{eq:eq:Tr_I_fin} and \eqref{part2} we have obtained that
\bea
   \label{eq:Fr-det}
    & -\Tr ( \mathcal{R}\circ \delta  \mathcal{K})
    = 
    -\Tr(\eqref{eq:I} + \eqref{eq:II})  =
    \nn\\
     &=-\iint_{\mathscr{D}}\Tr(\Gamma^{-1}(z)\partial_z\Gamma(z)\delta  (f(z)g^{T}(z)))\frac{\d\ov z \wedge \d{z}}{2i}=
     \nn\\
     &=-\iint_{\mathscr{D}}\Tr(\Gamma^{-1}(z)\partial_z\Gamma(z)\delta  M(z))\frac{\d\ov z \wedge \d{z}}{2\pi i}.
     \eea
This concludes the proof of Proposition~\ref{prop3.2}.     \QED
 \subsection{Malgrange one form and $\tau$-function}
\label{secMal}
 From  Proposition~\ref{prop3.2}  we  define the following one form on the space of deformations, which we call {\it Malgrange one form} following the terminology in \cite{Bert10}: 
\begin{equation}
  \label{eq:Malg_form}
  \omega:=-\iint_{\mathscr{D}}\Tr\Big(\Gamma^{-1}(z)\partial_z\Gamma(z)\delta M(z)\Big)\frac{\d \ov z\wedge \d {z}}{2 \pi i}, 
\end{equation}
where $\Gamma(z)$ is the solution of the $\ov\pa$-problem~\ref{dbarproblem} and $M(z)$ is defined in~\eqref{eq:M}.
For  the operator  $\mathcal K$ defined in \eqref{Koper}, the Proposition~\ref{prop3.2} implies that 
\bea
\omega = \delta \log \text{{\rm det}}_2(\id -\mathcal{K}), 
\eea
and hence $\omega$ is an exact (and hence closed) one form in the space of deformation parameters the operator $\mathcal K$ may depend upon. The form  $\omega$ can be shown to be closed under weaker assumptions on the matrix $M$ than the ones that appears in the $\ov\pa$-problem \ref{dbarproblem} as the following theorem shows.
\begin{theorem}
  \label{th_2}
Suppose that the $r \times r$  matrix $M = M(z,\ov z; \pmb{t})$ is smooth and compactly supported in $\mathscr{D}$  (uniformly with respect to the parameters $\pmb{t}$), depends smoothly on $\pmb{t}$ and   the matrix trace  $\Tr (M)\equiv 0$.
Let $\Gamma(z,\ov z;\pmb{t})$ be  the solution of the $\ov\pa$-problem \ref{dbar1}.  Then   the exterior differential of the one-form   $\omega $   defined in  \eqref{eq:Malg_form} vanishes:
\be
\delta \omega=0.
\ee
\end{theorem}

\noindent {\bf Proof.} 
From the $\ov\pa$-problem we obtain
\be
\label{deltaGamma}
\delta( \pa_{\bar{z}} \Gamma) = \Gamma \delta M + \delta \Gamma M \ \ \ \Rightarrow \ \ \ 
\delta \Gamma(z) = \iint_{\mathscr{D}}
   \frac{\Gamma(w)\delta M(w)\Gamma^{-1}(w)}{(w - z)^2}\frac{\d \ov w \wedge \d w}{2 \pi i}\Gamma(z).
\ee
Using \eqref{deltaGamma} we can compute 
\bea
\label{eq:proof}
\delta \omega &= -\iint_{\mathscr{D}}
   \Tr\Big(\delta(\Gamma^{-1}\partial_z\Gamma \wedge\delta  M )\Big)\frac{\d \ov z \wedge \d{z}}{2 \pi i} 
=\nn\\
&= \iint_{\mathscr{D}}
\Tr
\Big(\Gamma^{-1}\delta \Gamma \Gamma^{-1}\partial_z \Gamma \wedge\delta  M \Big)\frac{\d \ov z \wedge \d{z}}{2 \pi i}
-
 \iint_{\mathscr{D}}\Tr
 \Big(
 \Gamma^{-1}\delta \partial_z\Gamma \wedge \delta M\Big )\frac{\d \ov z \wedge \d{z}}{2 \pi i}
 \eea
 
From \eqref{deltaGamma} we deduce
\begin{equation}
  \label{eq:proof_G1}
  \delta \partial_z\Gamma(z) =
   -\iint_{\mathscr{D}}
   \frac{\Gamma(w)\delta M(w)\Gamma^{-1}(w)}{(w - z)^2}\frac{\d \ov w \wedge \d w}{2 \pi i}\Gamma(z) + \delta \Gamma(z) \Gamma(z)^{-1} \partial_z\Gamma(z).
\end{equation}
Substituting~\eqref{eq:proof_G1} in the equation~\eqref{eq:proof} we obtain:  

\begin{equation}
  \label{eq:proof_om}
\delta \omega = \iint_{\mathscr{D}}
 \Tr
 \left(
\Gamma^{-1}(z)\left(\iint_{\mathscr{D}}\frac{\Gamma(w)\delta  M(w)\Gamma^{-1}(w)}{(w - z)^{2}}\frac{\d \ov w \wedge d{w}}{2 \pi i}\right)\Gamma(z) \wedge \delta M(z)
\right)
\frac{\d \ov z \wedge \d  z}{2 \pi i}.
\end{equation}
The crux of the proof is now the correct evaluation of the iterated integral:
\bea
\delta \omega  &= \iint_{\mathscr{D}} \frac{ \d^2 z}\pi\iint_{\mathcal{\mathscr{D}}} \frac{ \d^2 w}\pi
\frac{ F(z,w)}{(z-w)^2},\nn \\
&  F(z,w):= \Tr
 \Big(
\Gamma(w)\delta  M(w)\Gamma^{-1}(w)  \wedge \Gamma(z)\delta M(z)\Gamma^{-1}(z)
\Big)
\eea
By applying Fubini's theorem, since the integrand is antisymmetric in the exchange of the variables $z\leftrightarrow w$, we quickly conclude that the integral is zero. However the integrand is singular along the diagonal $\Delta := \{ z=w\}\subset  \mathscr{D}\times \mathscr{D}$ and we need to make sure that the integrand is absolutely summable. 

Recalling that  $F(z,w)=-F(w,z)$, so that $F(z,z)\equiv 0$, we now compute the Taylor expansion of $F(z,w)$ with respect to $w$ near $z$;
\be
F(z,w) = 0+  \pa_{w}F(z,z) (w-z) + \pa_{\bar{w}}F(z,z) (\ov w- \ov z)  + \mathcal O(|z-w|^2).
\ee
Thus $\frac{|F(z,w)|}{|z-w|^2} = \mathcal O(|z-w|^{-1})$ which is integrable with respect to the area measure. Hence application of Fubini's theorem is justified. 
\QED
From this theorem, we can define a $\tau$-function associated to the the $\ov\pa$-problem~\ref{dbar1} by 
\begin{equation}
   \label{eq:tau}
   \tau(\pmb{t})= \exp\left(\int \omega\right). 
\end{equation}
 In general the  above  $\tau$-function is defined only up to scalar multiplication and hence should be rather thought of as a section of an appropriate line bundle over the space of deformation parameters, depending on the context. However, for $M$ in the form specified in \eqref{eq:M} we know from Proposition~\ref{prop3.2} that we can {\it identify} the $\tau$-function with the  regularized Hilbert-Carleman determinant: 
\begin{equation}
  \label{eq:taudet}
  \tau(\pmb{t})= \exp\left(\int \omega\right)= \text{{\rm det}}_2(\id -\mathcal{K}). 
\end{equation}
In the next section, by choosing a specific dependence on the parameters $\pmb{t}$  in the more general setting of $M$ as in Thm. \ref{th_2}  we are going to show that $ \tau(\pmb{t})$ is a  KP $\tau$-function in the sense that it satisfies Hirota bilinear relations \cite{Hirota1986}.

\section{$\tau (\pmb{t})$ as a KP $\tau$-function}
\label{sect:tau}

In this section we   consider a specific  type of dependence of $M$ on the ``times'': let  $M(z,\pmb{t})$   be a $2\times 2$ matrix that depends on $ \pmb{t}$ in  the following form
\begin{equation}
  \label{eq:M-matrix}
  M(z,\pmb{t})= {\rm e}^{\frac{\xi(z,\pmb{t})}{2}\sigma_{3}}M_{0}(z){\rm e}^{-\frac{\xi(z,\pmb{t})}{2}\sigma_{3}}\;,
\end{equation}
with \begin{equation}
\label{xi}
\xi(z,\pmb{t})= \sum_{j=1}^{+\infty}z^{j}t_{j}
\end{equation}
and $M_{0}(z,\bar{z})$ a traceless matrix compactly supported on $\mathscr{\mathscr{D}}$. 
A $\tau$-function of the Kadomtsev-Petviashvili  hierarchy, $\tau(\pmb{t})$,  can be characterized as a function of  (formally) an infinite number of variable which satisfies the Hirota Bilinear relation
\begin{equation}
  \label{eq:Hirota}
  \text{Res}_{z=\infty} (\tau(\pmb{t} -[z^{-1}])\tau(\pmb{s} + [z^{-1}]){\rm e}^{\xi(z,\pmb{t}) - \xi(z,\pmb{s})}=0
\end{equation}
where $\pmb{t} \pm [z^{-1}]$ is the \textit{Miwa Shift}, defined as:
\begin{equation}
  \label{eq:Miwa}
 \pmb{t} \pm [z^{-1}]:= \le(t_1 \pm \frac{1}{z}, t_2 \pm \frac{1}{2 z^{2}}, \dots, t_j \pm \frac{1}{j z^{j}}, \dots\ri).
\end{equation}
 The residue in \eqref{eq:Hirota} is meant in the formal sense, namely by considering the coefficient of $z^{-1}$ in the expansion at infinity and can be thought of as 
 the limit of $\oint_{|z|=R}$ as $R\to+\infty$.  If the functions of $z$  intervening  in \eqref{eq:Hirota} can be written as analytic functions in a deleted neighbourhood of  $\infty$, then the residue is  a genuine integral; this is the case of interest below.

 As described in~\cite{JMD}, the equation~\eqref{eq:Hirota} implies that the tau function satisfy an equation of the Hirota type
 \be
 P(\mathcal{D}_1, \mathcal{D}_2, \dots)\, \tau^2=0
 \ee
 where $\mathcal{D}_j$ is the Hirota derivative respect to $t_j$, defined as
 \be
 \mathcal{D}_j \, p(\pmb{t}) q(\pmb{t}) := (\pa_{t_j} - \pa_{t'_j})(p(\pmb{t}) q(\pmb{t}'))|_{\pmb{t}=\pmb{t}'},
 \ee
 and $P(\mathcal{D}_1, \mathcal{D}_2, \dots)$ is a polynomial in $(\mathcal{D}_1, \mathcal{D}_2, \dots)$. In particular, if we consider the first three times $t_1,t_2 \text{ and } t_3$,  and $t_k=0$ for $k\geq 3$ the equation~\eqref{eq:Hirota} is equivalent to the KP equation in Hirota's form
 \be
 ( 3 \mathcal{D}^2_2 -4 \mathcal{D}_1 \mathcal{D}_3 +\mathcal{D}^{4}_1) \, \tau^2 =0.
 \ee
 Putting 
 \[
   \pa^2_{t_1}\log \tau(t_1,t_2,t_3)=\frac{1}{2}u(t_1,t_2,t_3)
 \]
 one obtains the celebrated KP equation
   \begin{equation}
   \label{KP}
    3   \partial^2_{t_2}u =   \partial_{t_1}(4  \partial_{t_3} u  -   \partial^3_{t_1}  u  - 6 u  \partial_{t_1} u )
    \end{equation}

The rest of this section is devoted to the  verification  of   the Hirota bilinear relation \eqref{eq:Hirota}  for the  KP tau function.
\subsection{Hirota bilinear relation for  the KP hierarchy}
The main result is the following.
\begin{theorem}
    \label{th-tau}
    Let $\Gamma(z,\pmb{t})$ be the solution of the $\ov\pa$-problem 
    \[
    \pa_{\bar{z}} \Gamma(z,\pmb{t}) = \Gamma(z,\pmb{t}) {\rm e}^{\frac{\xi(z,\pmb{t})}{2}\sigma_{3}}M_{0}(z){\rm e}^{-\frac{\xi(z,\pmb{t})}{2}\sigma_{3}},\;\qquad \Gamma(z,\pmb{t}) \underset{z \to \infty}{ \to } \1
    \] 
    with the  traceless   matrix $M_0(z)$   compactly supported on   a bounded domain  $\mathscr{D}$ of the complex plane and the function 
    $\xi$ given by the formal sum $\xi(z,\pmb{t})= \sum_{j=1}^{+\infty}z^{j}t_{j}$.
    Then the  function 
    \begin{equation}
    \label{KPtau}
    \tau(\pmb{t})=\exp\left(\int\omega\right),
    \end{equation}
    with $\omega$  defined in \eqref{eq:Malg_form}
     is a KP $\tau$-function; i.e. it satisfies the Hirota Bilinear relation~\eqref{eq:Hirota}.
  \end{theorem}
 
\begin{remark}  In this setting the KP $\tau$-function  is in general complex--valued.  Under appropriate additional symmetry constraints for the matrix $M_0$   and the domain $\mathscr{D}$ we can obtain  a real--valued $\tau$-function.
\end{remark}

We prove the theorem in several steps. We first analyse the effect of the Miwa shifts on the $\tau$-function.
For this purpose we need to determine how the Miwa shift acts on the matrices $\Gamma(z,\bar{z},t)$ and $M(z,\bar{z},\pmb{t})$. We consider $M(z,\bar{z},\pmb{t} \pm [\zeta^{-1}])$ first. 
\begin{equation*}
   M(z,\pmb{t} \pm [\zeta^{-1}])= {\rm e}^{\xi(z,\pmb{t}\pm [\zeta^{-1}])\sigma_{3}}M_{0}(z){\rm e}^{-\xi(z,\pmb{t} \pm [\zeta^{-1}])\sigma_{3}}
 \end{equation*}
 from the definition of $\xi(z,t)$~\eqref{xi}
\begin{equation*}
  \begin{split}
   &\xi(z,\pmb{t} \pm [\zeta^{-1}])= \sum^{+\infty}_{j=1}z^{j}\left(t_{j} \pm \frac{1}{j \zeta^{j}} \right)= \sum^{+\infty}_{j=1} z^{j}t_j \pm \sum^{+\infty}_{j=1}\frac{z^{j}}{j \zeta^{j}}= \xi(z,\pmb{t}) \mp \ln\left( 1 - \frac{z}{\zeta}\right)\\
  \end{split}
\end{equation*}
 and we have that
\begin{equation}
  \label{eq:M_s}
  M(z,\pmb{t} \pm [\zeta^{-1}])=\left(1 - \frac{z}{\zeta}\right)^{\mp \frac{\sigma_{3}}{2}}M(z,\pmb{t})\left(1 - \frac{z}{\zeta}\right)^{\pm \frac{\sigma_{3}}{2}}.
\end{equation}
For the matrices $\Gamma(z,\ov z,\pmb{t} \pm [\zeta^{-1}])$ we need to consider the two case separately. Let us start with the negative shift $\Gamma(z,\ov z,\pmb{t} -[\zeta^{-1}])$.
\begin{equation}
  \label{eq:G-shift}
  \begin{split}
    \partial_{\bar{z}}\Gamma(z,\pmb{t} -[\zeta^{-1}])&=\Gamma(z,\pmb{t} -[\zeta^{-1}])M(z,\pmb{t} - [\zeta^{-1}])\\
    &=\Gamma(z,\pmb{t} -[\zeta^{-1}])\left(1 - \frac{z}{\zeta}\right)^{+ \frac{\sigma_{3}}{2}}M(z,\pmb{t})\left(1 - \frac{z}{\zeta}\right)^{- \frac{\sigma_{3}}{2}}\\
    &=\Gamma(z,\pmb{t} -[\zeta^{-1}])D(z,\zeta)M(z,\pmb{t})D^{-1}(z,\zeta)\\
  \end{split}
\end{equation}
where
\be
D(z,\zeta)=
    \begin{bmatrix}
      1 - \frac{z}{\zeta} & 0\\
      0 & 1
    \end{bmatrix}. \nn
\ee
From~\eqref{eq:G-shift}, we notice that the matrix $\Gamma(z,t - [\zeta^{-1}])D(z,\zeta)$ satisfies the $\ov\pa$-problem~\ref{dbarproblem}, i.e. there exists a connection matrix $C(z)$ such that
\begin{equation}
  \label{eq:Gamma-1}
  \Gamma(z,\pmb{t} -[\zeta^{-1}])= C(z)\Gamma(z,\pmb{t})D(z,\zeta)^{-1},
\end{equation}
where obviously $C(z)$ depends also on $\zeta$ and $\pmb{t}$.

 The matrix $C(z)$ is determined by the conditions that both $\Gamma(z,\pmb{t})$ and $\Gamma(z,\ov z, \pmb{t}-[\zeta^{-1}])$ must tend to $\1$ for $z\to \infty$ and are regular at $z=\zeta$
\begin{equation}
  \label{eq:R-cond}
  \begin{split}
    \lim_{z\to \infty}\left(1 - \frac{z}{\zeta}\right)^{-1}&C(z)
    \begin{bmatrix}
      \Gamma_{1 1}(z,\pmb{t})\\
      \Gamma_{1 2}(z,\pmb{t})
    \end{bmatrix}=
    \begin{bmatrix}
      1\\
      0
    \end{bmatrix}
\quad  \lim_{z\to \infty}C(z)
    \begin{bmatrix}
     \Gamma_{2 1}(z,\pmb{t})\\
      \Gamma_{2 2}(z,\pmb{t})
    \end{bmatrix}=
    \begin{bmatrix}
      0\\
      1
    \end{bmatrix}\\
    &\lim_{z \to \zeta} \left(1 - \frac{z}{\zeta}\right)^{-1}C(z)
    \begin{bmatrix}
      \Gamma_{1 1}(z,\pmb{t})\\
      \Gamma_{1 2}(z,\pmb{t})
    \end{bmatrix}=
    \begin{bmatrix}
      \Gamma_{1 1}(\zeta,\pmb{t})\\
      0
    \end{bmatrix}
  \end{split}
\end{equation}
Solving the  system~\eqref{eq:R-cond}, we obtain that the matrix $C(z)$ has the following form
\begin{equation}
  \label{eq:R-mat}
  C(z)=
  \begin{bmatrix}
    \left(1 - \frac{z}{\zeta}\right) + \frac{\partial_z\Gamma_{12}(\infty)\Gamma_{21}(\zeta)}{\zeta \Gamma_{11}(\zeta)} & -\frac{\partial_z\Gamma_{12}(\infty)}{\zeta}\\
    -\frac{\Gamma_{21}(\zeta)}{\Gamma_{11}(\zeta)} & 1
  \end{bmatrix}.
\end{equation}

Following the same ideas, we can find a similar formula for $\Gamma(z,\ov z,t + [\zeta^{-1}])$
\begin{equation}
  \label{eq:Gamma-2}
  \Gamma(z,\pmb{t} + [\zeta^{-1}])= \tilde{C}(z)\Gamma(z,\pmb{t})\tilde{D}(z,\zeta)^{-1}
\end{equation}
with
\begin{equation*}
  \tilde{D}(z,\zeta)=
    \begin{bmatrix}
      1 & 0\\
      0 & 1 - \frac{z}{\zeta}
    \end{bmatrix}.
  \end{equation*}
  Also in this case, we have three conditions similar to~\eqref{eq:R-cond} :
  \begin{equation}
  \label{eq:R-cond2}
  \begin{split}
    \lim_{z\to \infty}\tilde{C}(z)
    &\begin{bmatrix}
      \Gamma_{1 1}(x,\pmb{t})\\
      \Gamma_{1 2}(x,\pmb{t})
    \end{bmatrix}=
    \begin{bmatrix}
      1\\
      0
    \end{bmatrix}
\quad  \lim_{z\to \infty} \left(1 - \frac{z}{\zeta}\right)^{-1}\tilde{R}(z,\zeta)
    \begin{bmatrix}
      \Gamma_{2 1}(x,\pmb{t})\\
      \Gamma_{2 2}(x,\pmb{t})
    \end{bmatrix}=
    \begin{bmatrix}
      0\\
      1
    \end{bmatrix}\\
    &\lim_{z \to x} \left(1 - \frac{z}{\zeta}\right)^{-1}\tilde{C}(z)
    \begin{bmatrix}
      \Gamma_{2 1}(x,\pmb{t})\\
      \Gamma_{2 2}(x,\pmb{t})
    \end{bmatrix}=
    \begin{bmatrix}
      0\\
      \Gamma_{22}(\zeta,\pmb{t})
    \end{bmatrix}
  \end{split}
\end{equation}
and we find out that $\tilde{C}(z)$ has the following form:
\begin{equation}
  \label{eq:R-mat2}
 \tilde{C}(z)=
  \begin{bmatrix}
    1  & -\frac{\Gamma_{12}(\zeta)}{\Gamma_{22}(\zeta)}\\
    -\frac{\partial_z\Gamma_{21}(\infty)}{\zeta} & \left(1 - \frac{z}{\zeta}\right) + \frac{\partial_z\Gamma_{21}(\infty)\Gamma_{12}(\zeta)}{\zeta \Gamma_{22}(\zeta)}
  \end{bmatrix}.
\end{equation}

We need to show how the Miwa shift acts on the Malgrange one form. We define $\delta_{[\zeta]}$ the differential deformed including the external parameter $\zeta$
\begin{equation}
  \label{eq:def-delt}
    \delta_{[\zeta]} := \sum^{+\infty}_{j=1} \d t_{j} \partial_{t_{j}} + \d \zeta \partial_{\zeta}= \delta + \delta_{\zeta}.
  \end{equation}

\begin{lemma}
  \label{lemm-2}
  When  $\zeta \notin \mathscr{D}$   the Miwa shift~\eqref{eq:Miwa} acts on the Malgrange one form~\eqref{eq:Malg_form} in the following way:
  \be
  \label{eq:lemm-2}
 \omega(\pmb{t} \pm [\zeta^{-1}])= \omega(\pmb{t}) + \delta_{[\zeta]}\ln\le((\Gamma^{\mp 1}(\zeta))_{11}\ri) \mp \delta_{[\zeta]}\gamma(\zeta)\,,
  \ee
  where $\Gamma(z)$ solves the $\ov\pa$-problem \ref{dbarproblem} and $\gamma(\zeta)$ is a $\pmb{t}$ independent function defined as
  \be
  \label{eq:gam}
  \gamma(\zeta):= \iint_{\mathscr{D}}\log\left(\frac{\zeta}{\zeta-z}\right)(\partial_z M_{0}(z))_{11}\frac{ \d \ov{z} \wedge \d z}{2\pi i}\,,\;\zeta\in\C\backslash \mathscr{D},
 \ee
that is  is analytic  (for $\zeta \notin \mathscr{D}$) and goes to zero as $\zeta \to \infty$.
\end{lemma}
Observe that since $\Tr M_0=0$ we may express the formula in terms of the $(2,2)$ entry instead. 
The proof of this lemma is presented in the Appendix~\ref{app-II}.
Now we can state the following proposition:

\begin{proposition}
  For  $\zeta \notin \mathscr{D}$  the following relations holds: 
  \begin{equation}
    \label{eq:tau-M}
    \frac{\tau(\pmb{t} -[\zeta^{-1}])}{\tau(\pmb{t})} = \Gamma_{11}(\zeta,\pmb{t}){\rm e}^{\gamma(\zeta)}  \qquad \frac{\tau(\pmb{t} + [\zeta^{-1}])}{\tau(\pmb{t})} = \Gamma^{-1}_{11}(\zeta,\pmb{t}){\rm e}^{-\gamma(\zeta)}\,,
  \end{equation}
  where $\tau(t)$ is defined in~\eqref{eq:tau},
  $\Gamma(z)$ solves the $\ov\pa$-problem \ref{dbarproblem} and $\gamma(\zeta)$ is defined in~\eqref{eq:gam}
\end{proposition}

\noindent {\bf Proof.}
    From Lemma~\ref{lemm-2} and the equation~\eqref{eq:tau}, we  rewrite~\eqref{eq:lemm-2} as 
    \begin{equation}
      \label{eq:lnt}
      \delta_{[\zeta]} \ln \tau(\pmb{t} \pm [\zeta^{-1}])= \delta_{[\zeta]} \ln\tau(\pmb{t}) + \delta_{[\zeta]}\ln((\Gamma^{\mp 1}(\zeta))_{11}) \mp \delta_{[\zeta]}\gamma(\zeta)
    \end{equation}
    an then, from the properties of the  logarithm the statement~\eqref{eq:tau-M} is proved.
  \QED

\begin{remark}
  The exponential term ${\rm e}^{\gamma(\zeta)}$ could be absorbed by a gauge transformation in the formalism of the infinite dimensional Grassmannian manifold of Segal-Wilson (\cite{Harnad2021}, Chapter 4). Such  gauge transformations have no effect on the Hirota bilinear relation~\eqref{eq:Hirota}.
\end{remark}

  
 Let us now  define the matrix $H(z)$ as
  \begin{equation}
      \label{eq:lemm-1}
      H(z):=H(z; \pmb{t}, \pmb{s}):= \Gamma(z,\pmb{t}){\rm e}^{(\xi(z,\pmb{t}) -\xi(z,\pmb{s}))E_{11}}\Gamma^{-1}(z,\pmb{s})
  \end{equation}
  where $E_{11}= \begin{bmatrix} 1 & 0\\ 0 & 0\end{bmatrix}$, $\Gamma(z,\ov z, \pmb{t})$ solves the $\ov\pa$-problem~\ref{dbarproblem} and $\pmb{s}= (s_1, s_2, \dots, s_j,\dots)$ denotes another set of values for the deformation parameters. 
  
  \begin{lemma}
    \label{lemma-1}
    
    The matrix $H(z)$ defined in~\eqref{eq:lemm-1} is analytic for all $z \in \mathbb{C}$.
  \end{lemma}

  \noindent {\bf Proof.}
    For $z \notin \mathscr{D}$ the statement is trivial, so we consider the case of $z \in \mathscr{D}$.
    
    We apply the operator $\pa_{\bar{z}}$ to the matrix~\eqref{eq:lemm-1}
    \begin{equation*}
      \begin{split}
        \pa_{\bar{z}} H(z)&=\pa_{\bar{z}}\Gamma(z,\pmb{t}){\rm e}^{(\xi(z,\pmb{t}) -\xi(z,\pmb{s}))E_{11}}\Gamma^{-1}(z,\pmb{s}) +\Gamma(z,\pmb{t}){\rm e}^{(\xi(z,\pmb{t}) -\xi(z,\pmb{s}))E_{11}}\pa_{\bar{z}}\Gamma^{-1}(z,\pmb{s})\\
        &= \Gamma(z,\pmb{t})M(z,\pmb{t}){\rm e}^{(\xi(z,\pmb{t}) -\xi(z,\pmb{s}))E_{11}}\Gamma^{-1}(z,\pmb{s})  +\\
        &\quad -\Gamma(z,\pmb{t}){\rm e}^{(\xi(z,\pmb{t}) -\xi(z,\pmb{s}))E_{11}}M(z,\pmb{s})\Gamma^{-1}(z,\pmb{s})\\
        &=\Gamma(z,\pmb{t})\left({\rm e}^{\frac{\xi(z,\pmb{t})}{2}\sigma_{3}}M_{0}(z){\rm e}^{-\frac{\xi(z,\pmb{t})}{2}\sigma_{3}}{\rm e}^{(\xi(z,\pmb{t}) -\xi(z,\pmb{s}))E_{11}} +\right. \\
        &\quad \left.-{\rm e}^{(\xi(z,\pmb{t}) -\xi(z,\pmb{s}))E_{11}}{\rm e}^{\frac{\xi(z,\pmb{s})}{2}\sigma_{3}}M_{0}(z){\rm e}^{-\frac{\xi(z,\pmb{s})}{2}\sigma_{3}}\right)\Gamma^{-1}(z,\pmb{s})\\
        &=\Gamma(z,\pmb{t})\left({\rm e}^{\frac{\xi(z,\pmb{t})}{2}\sigma_{3}}{\rm e}^{\frac{\xi(z,\pmb{t})}{2}\mathbb{I}}M_{0}(z){\rm e}^{-\xi(z,\pmb{s})E_{11}} +\right. \\
        &\quad \left. -{\rm e}^{\xi(z,\pmb{t})E_{11}}M_{0}(z){\rm e}^{\frac{\xi(z,\pmb{s})}{2}\mathbb{I}}{\rm e}^{-\frac{\xi(z,\pmb{s})}{2}\sigma_{3}}\right)\Gamma^{-1}(z,\pmb{s})\\
        &=\Gamma(z,\pmb{t})\left({\rm e}^{\xi(z,\pmb{t})E_{11}}M_{0}(z){\rm e}^{-\xi(z,\pmb{s})E_{11}} -{\rm e}^{\xi(z,\pmb{t})E_{11}}M_{0}(z){\rm e}^{-\xi(z,\pmb{s})E_{11}} \right)\Gamma^{-1}(z,\pmb{s})\\
        &= 0
      \end{split}
    \end{equation*}
    and this proves the statement.
  \QED

  We are now ready to  prove the main result of the section,  namely  Theorem~\ref{th-tau}.

    \noindent {\bf Proof   of Theorem~\ref{th-tau}.}
      Let us compute the residue
      \bea
      \label{eq:proof-hir}
         & \text{Res}_{z=\infty}(\tau(\pmb{t} -[z^{-1}])\tau(\pmb{s} + [z^{-1}]){\rm e}^{\xi(z,\pmb{t}) - \xi(z,\pmb{s})}) \nn\\
         & = \tau(\pmb{t})\tau(\pmb{s})\text{Res}_{z=\infty}(\frac{\tau(\pmb{t} -[z^{-1}])}{\tau(\pmb{t})}\frac{\tau(\pmb{s} + [z^{-1}])}{\tau(\pmb{s})}{\rm e}^{\xi(z,\pmb{t}) - \xi(z,\pmb{s})})\nn\\
         & =\tau(\pmb{t})\tau(\pmb{s})\text{Res}_{z=\infty}(\Gamma_{11}(z,\pmb{t})(\Gamma^{-1}(z,\pmb{s}))_{11}{\rm e}^{\xi(z,\pmb{t}) - \xi(z,\pmb{s})}) \nn\\
         & = \tau(\pmb{t})\tau(\pmb{s})\lim_{R\to \infty}\oint_{|z|=R}\Gamma_{11}(z,\pmb{t}){\rm e}^{\xi(z,\pmb{t}) - \xi(z,\pmb{s})}(\Gamma^{-1}(z,\pmb{s}))_{11}\frac{\d z}{2\pi i}
        \eea
        Consider the first diagonal element of the matrix $H(z)$. From the analyticity of $H(z)$ proved in Lemma~\ref{lemma-1} we get 
        \begin{equation}
          \begin{split}
       0= \left(\oint_{|z|=R} H(z)\frac{\d z}{2\pi i}\right)_{11} &=\oint_{|z|=R}\Gamma_{11}(z,\pmb{t}){\rm e}^{\xi(z,\pmb{t}) -\xi(z,\pmb{s})}(\Gamma^{-1}(z,\pmb{s}))_{11}\frac{\d z}{2\pi i} +\\
        &\quad - \oint_{|z|=R}\Gamma_{12}(z,\pmb{t})\Gamma_{21}(z,\pmb{s})\frac{\d z}{2\pi i}.
        \end{split}
        \end{equation}
        So, we can rewite~\eqref{eq:proof-hir} as
        \be
        \label{eq:proof-hir-2}
        ~\eqref{eq:proof-hir}=\tau(\pmb{t})\tau(\pmb{s})\lim_{R\to \infty}\oint_{|z|=R}\Gamma_{12}(z,\pmb{t})\Gamma_{21}(z,\pmb{s})\frac{\d z}{2\pi i}
        \label{eerw}
        \ee
         Since both $\Gamma_{12}(z,\pmb{t})$ and $\Gamma_{21}(z,\pmb{s})$ are analytic for $|z|$ sufficiently large (given that $\mathscr D$ is compact) and
      \begin{equation*}
        \Gamma(z,\pmb{t}) \sim \mathbb{I} + \mathcal{O}(z^{-1})\quad \mbox{for $z \to \infty$},
      \end{equation*}
      it follows that  \eqref{eerw} is zero because the integrand is $\mathcal O(z^{-2})$, and the statement is proved.
    \QED%
    \subsection{The  focusing Nonlinear Schr\"odinger equation}
    \label{NonSCH}
    In this subsection we  make a specific choice of the matrix $M_0$   of the form
         \[
     M_0(z)=\begin{bmatrix}
     0&\beta(z)\chi_{\mathscr{D}}\\
     -\overline{\beta(\overline{z})}\chi_{\overline{\mathscr{D}}}&0
     \end{bmatrix},
     \]
     where $\beta(z)= \beta(z,\ov z)$ is a smooth function on $\mathscr{D}\subset\C_+$  and $\chi_{\mathscr{D}}$  ($\chi_{\overline{\mathscr{D}}}$) is the characteristic function of       $\mathscr{D}$   ($\overline{\mathscr{D}}$).
     We observe that $M_0$  satisfies the Schwarz symmetry
     \be
     \label{eq:Sch_sym}
      \overline{M_0(\ov z)}=\sigma_2 M_0(z)\sigma_2 , \quad \text{ where } \sigma_2=
      \begin{bmatrix}
        0 & -i\\
        i & 0
      \end{bmatrix}.
      \ee
Let us consider the $\ov\pa$-problem
        \bea
     \label{eq:NLShdbar}
     &  \pa_{\bar{z}} \Gamma(z,\pmb{t}) = \Gamma(z,\pmb{t}){\rm e}^{-i\xi(z,\pmb{t})\sigma_3} M_0(z){\rm e}^{i\xi(z,\pmb{t})\sigma_3} & \text{for } z \in {\mathscr D} \cup \ov  {\mathscr D}  \\
   &  \Gamma(z,\pmb{t}) \underset{z \to \infty}{ \to } \1 & \nn 
     \eea
     with  $\xi(z,\pmb{t})$ as in \eqref{xi}.  Here we have re-defined  $t_j\to -2i t_j$ to respect  the customary  normalization of times in the KP-hieararchy.
     
 \begin{theorem}
      \label{th-NLSh}
       Let  $\Gamma(z, \pmb{t})$ be the solution of the $\ov\pa$-problem~\eqref{eq:NLShdbar}  and let
       \[
      \psi(\pmb{t}):=2i\lim_{z\to\infty}z(\Gamma(z, \pmb{t})-\1)_{12}\,.
       \]
       Then the function $\psi=\psi(\pmb{t})$  satisfies the nonlinear Schr\"odinger hierarchy~\cite{FadTak,Matveev2018}  written in the recursive form
        \bea
        & i\pa_{t_m} \psi_1 = 2 \psi_{m+1},\quad      \psi_1:=\psi \label{eq:thi}, \;\;m\geq 1,\\
        & \psi_{m} = \frac{i}{2}\pa_{t_1} \psi_{m-1} + \psi_1  h_{m-1}, \quad \pa_{t_1} h_{m}= 2 \im(\psi_1 \ov \psi_{m}),
          \label{eq:thf}
        \eea
        where $\psi_{m}$ and $h_{m}$ are functions of $\pmb{t}$ and $h_1:=0$.
      \end{theorem}
      The proof of this theorem is classical and  is deferred to  Appendix~\ref{app:I}.
      In particular the second flow gives the focusing NLS equation
\[
        i \pa_{t_2}   \psi + \frac{1}{2} \pa^{2}_{t_1} \psi + |\psi|^{2}\psi=0,
  \]    
  where comparing with the notation in the  introduction $t_2=t$ and $t_1=x$.  The third flows gives the so called complex modified KdV equation
   \be
    \pa_{t_3} \psi + \frac{\pa^{3}_{t_1} \psi}{4} + \frac{3}{2}|\psi|^2\pa_{t_1} \psi=0. \nn
    \ee
  Setting $t_k=0$ for $k\geq 4$ one obtains that $v(t_1,t_2,t_3):=2|\psi_1(t_1,t_2,t_3)|^2$ satisfies  the KP equation 
  \eqref{KP} after the  rescalings   $v=-4u$ and $t_j\to \frac{i}{2}t_j$.
\subsubsection{Reduction to a Riemann-Hilbert problem}
    In this section we show that for particular choices of $\beta$ and of the domain $\mathscr{D}$ one can reduce the $\ov\pa$-problem \eqref{eq:NLShdbar}  to a standard Riemann-Hilbert problem when $z$ is  outside $\mathscr{D}$ and $\ov{\mathscr{D}}$.
   The easiest way to solve the $\ov\pa$-problem \eqref{eq:NLShdbar}  is to split it in components
\begin{equation*}
  \Gamma(z,\pmb{t})=[\vec{A}(z,\pmb{t}) \quad \vec{B}(z,\pmb{t})]
\end{equation*}
so that 
\begin{equation}
  \label{eq:d_bar_1}
  \begin{split}
    \bar{\partial} \vec{A}(z,\pmb{t}) &= 0\\
    \bar{\partial} \vec{B}(z,\pmb{t}) &= \beta(z)e^{-2i\xi(z,\pmb{t})}\vec{A}(z,\pmb{t})
  \end{split} \text{\quad for } z \in \mathcal{D}
\end{equation}

\begin{equation}
  \label{eq:d_bar_2}
  \begin{split}
    \bar{\partial} \vec{A}(z,\pmb{t}) &= -\beta^{*}(z)e^{2i\xi(z;\pmb{t})}\vec{B}(z,\pmb{t})\\
    \bar{\partial} \vec{B}(z,\pmb{t}) &= 0
  \end{split} \text{\quad for } z \in \mathcal{\overline{D}}
\end{equation}

with the boundary condition
\begin{equation}
  \vec{A} \sim
  \begin{pmatrix}
    1\\
    0
  \end{pmatrix} \quad
  \vec{B} \sim
  \begin{pmatrix}
    0\\
    1
  \end{pmatrix} \quad  \text{ as } z \to \infty .
\end{equation}

From the  equations~\eqref{eq:d_bar_1} and~\eqref{eq:d_bar_2}  one deduces that \begin{itemize}
\item $\vec{A}(z,\pmb{t})$ is analytic for $z$ in the set $\mathscr{D}$;
\item $\vec{B}(z,\pmb{t})$ is analytic for $z$ in the set $\mathscr{\overline{D}}$.  
\end{itemize}
From the Cauchy-Pompeiu formula, we can rewrite the equations~\eqref{eq:d_bar_1} and~\eqref{eq:d_bar_2} as a system of two integral equations
\begin{equation}
  \label{eq:int_prob}
  \begin{split}
    \vec{A}(z,\pmb{t}) &=
    \begin{bmatrix}
      1\\
      0
    \end{bmatrix} + \iint_{\mathscr{\overline{D}}}\frac{\vec{B}(w,\pmb{t})\beta^{*}(w)e^{2i\xi(z,\pmb{t})}}{w - z} \frac{d\bar{w} \wedge dw}{2 \pi i}\\
    \vec{B}(z,\pmb{t}) &=
    \begin{bmatrix}
      0\\
      1
    \end{bmatrix} - \iint_{\mathscr{D}}\frac{\vec{A}(w,\pmb{t})\beta(w)e^{-2i\xi(z,\pmb{t})}}{w -z} \frac{d\bar{w} \wedge dw}{2 \pi i}.\quad 
  \end{split}
\end{equation}
Let us assume now that $\beta(z)$ is analytic in $\mathscr{D}$  simply connected and the  boundary of  $\mathscr{D}$ is sufficiently smooth so that it can be  described by the so--called   Schwarz function  $S(z)$ \cite{Gustafsson}  of the domain $\mathscr{D}$  through the equation
\[
\overline{z}=S(z).
\]
 The Schwarz function admits  an  analytic extension to  a maximal domain $\mathscr D^0\subset \mathscr D$.  Here we assume  that $\mathscr{L}:=\mathscr D\setminus \mathscr D^0$  consist of a {\it mother-body}, i.e.,  a collection of smooth arcs. An example of this is the ellipse.
 Using Stokes theorem and the Schwarz function of the domain,  we can reduce the area integral in \eqref{eq:int_prob} to a contour integral, namley
 \begin{equation}
  \label{eq:int_prob1}
  \begin{split}
    \vec{A}(z,\pmb{t}) &=
    \begin{bmatrix}
      1\\
      0
    \end{bmatrix} + \oint_{\partial\mathscr{\overline{D}}}\frac{\vec{B}(w,\pmb{t})S^*(w)\beta^{*}(w)e^{2i\xi(z,\pmb{t})}}{w - z} \frac{d\bar{w} \wedge dw}{2 \pi i}\\
    \vec{B}(z,\pmb{t}) &=
    \begin{bmatrix}
      0\\
      1
    \end{bmatrix} - \oint_{\partial\mathscr{D}}\frac{\vec{A}(w,\pmb{t})S(w)\beta(w)e^{-2i\xi(z,\pmb{t})}}{w -z} \frac{d\bar{w} \wedge dw}{2 \pi i},\;z\notin \mathscr{D},
  \end{split}
\end{equation}
where the boundary $\partial \mathscr{D}$ is oriented anticlockwise.  By analiticity, we can shrink the contour integral to the mother-body $\mathscr{L}$, namely
 \begin{equation}
  \label{eq:int_prob2}
  \begin{split}
    \vec{A}(z,\pmb{t}) &=
    \begin{bmatrix}
      1\\
      0
    \end{bmatrix} + \oint_{\mathscr{\overline{L}}}\frac{\vec{B}(w,\pmb{t})\Delta S^*(w)\beta^{*}(w)e^{2i\xi(z,\pmb{t})}}{w - z} \frac{d\bar{w} \wedge dw}{2 \pi i}\\
    \vec{B}(z,\pmb{t}) &=
    \begin{bmatrix}
      0\\
      1
    \end{bmatrix} - \oint_{\mathscr{L}}\frac{\vec{A}(w,\pmb{t})\Delta S(w)\beta(w)e^{-2i\xi(z,\pmb{t})}}{w -z} \frac{d\bar{w} \wedge dw}{2 \pi i},\;z\notin \mathscr{D},
  \end{split}
\end{equation}
where $\Delta S(w)=S_-(w)-S_+(w)$,  with $S_{\pm}(z)$  the boundary values of $S$ on the oriented contour $\mathscr{L}$.
The orientation of $\mathscr{L}$ is inherited by the orientation of $\partial \mathscr{D}$.
We can express the system \eqref{eq:int_prob2}  in matrix form 
\begin{equation}
  \widetilde\Gamma(z,\pmb{t})= \1 + \int_{\mathscr{L}\cup\ov{\mathscr{L}}}\frac{\widetilde\Gamma(w,\pmb{t})\e^{-i\xi(z,\pmb{t})\sigma_3}\widetilde{M}(w,\pmb{t})\e^{i\xi(z,\pmb{t})\sigma_3}}{w - z}\frac{dw}{2 \pi i}
\end{equation}

where
\begin{equation}
 \widetilde{M}(z,\pmb{t})=
  \begin{bmatrix}
    0 & \Delta S(z) \beta(z)\chi_{\mathscr{L}}(z)\\
    -\Delta S^*(z) \beta^{*}(z)\chi_{\ov{\mathscr{L}}}(z)  & 0
  \end{bmatrix}
\end{equation}
Using then the Sokhotski-Plemelj formula we can rewrite the above integral equation  as a Riemann-Hilbert problem  for a matrix function $ \widetilde\Gamma(z,\pmb{t})$ analytic in $\C\backslash  \{\mathscr{L}\cup\ov{\mathscr{L}}\}$ such that 
\begin{equation}
\begin{split}
  \label{eq:ell_R_H}
  \widetilde\Gamma_{+}(z,\pmb{t})&=\widetilde\Gamma_{-}(z)e^{-i\xi(z,\pmb{t})\sigma_3}(\1+\widetilde{M}(z,\pmb{t}))\e^{i\xi(z,\pmb{t})\sigma_3},\quad z\in \mathscr{L}\cup\ov{\mathscr{L}},\\
  \widetilde\Gamma(z,\pmb{t})&=\1+\mathcal{O}(z^{-1}),\quad \mbox{as $z\to\infty$}.
  \end{split}
\end{equation}
We remark that $\Gamma(z,\pmb{t})$ and $  \widetilde\Gamma(z,\pmb{t})$  coincides only for $z\in\C\backslash\{\mathscr{D}\cup\ov{\mathscr{D}}\}$.  However for our purpose, namely the solution of the nonlinear Scr\"odinger equation, only the terms of
$\widetilde\Gamma(z,\pmb{t})$ for $z\to\infty$ are needed.  When the domain $\mathscr{D}$ is an ellipse we show in \cite{BGO2023} that the initial data for the nonlinear Scr\"odinger equation is   step-like   oscillatory.

\color{black}

    \section{Conclusions}
     
    The $\ov\pa$-problems treated in this manuscript differ from the $\ov\pa$  introduced in \cite{MM},\cite{MDM} to study asymptotic behaviour of orthogonal polynomials or PDEs with non analytic initial data respectively.
    In those cases the $\ov\pa$-problem  is a by-product of the steepest descent Deift-Zhou method extended to the case where the jump-matrix is not analytic but otherwise the initial problem is an ordinary RHP;   in our case, the 
    initial data is defined from the solution of the $\ov\pa$-problem and is encoded in the domain $\mathscr{D}$ 
    and in the matrix $M$  of the $\ov\pa$-problem \eqref{dbar0}.
    An equation similar to~\eqref{eq:NLShdbar} was also studied by Zhu et al.~\cite{ZJW}, with the aim to find solutions for the defocusing/focusing NLS with nonzero boundary conditions.

    A generalization that could be considered is one where instead of the ``pure'' $\ov\pa$-problem \eqref{dbarproblem} one has a mixed $\ov\pa$ and Riemann--Hilbert problem; this would correspond to an operator for example acting on $L^2(\mathscr D, \d^2z) \oplus L^2(\Sigma, |\d z|)$ (typically with $\pa \mathscr D\subseteq \Sigma$); this type of problems would use, in the computation of the exterior derivative of the Malgrange form, the full Cauchy--Pompeiu formula. We defer this investigation to future efforts.    
    
 \vspace{1cm}  
\noindent {\bf Acknowledgements}

\noindent
 TG  and GO acknowledge the support of   the European Union's H2020 research and innovation programme under the Marie 
 Sklodowska-Curie grant No. 778010 {\it   IPaDEGAN}, the GNFM-INDAM group and the  research project Mathematical Methods in NonLinear Physics (MMNLP), Gruppo 4-Fisica Teorica of INFN.  
The work of MB was supported in part by the Natural Sciences and Engineering Research Council of Canada (NSERC) grant RGPIN-2023-04747.

 \appendix
 \section{Connection between the $\ov\pa$-Problem and the Inverse Scattering Theory}
 \label{app:I}
In this section we   prove Theorem~\ref{th-NLSh}  by deriving the corresponding   Zakharov-Shabat Lax pair \cite{ZS} for the solution  of   the $\ov\pa$-problem~\eqref{eq:NLShdbar}.
 To simplify the presentation, we restrict only to the first flow, namely we set $t_1=x$, $t_2=t$ and $t_j=0$ for $j\geq 3$.  The general case can be treated in a similar way.
Let us consider the matrix
\begin{equation}
  \label{eq:psi_1}
  \Psi(z;x,t)=\Gamma(z;x,t){\rm e}^{-i(zx  +z^2 t)\sigma_{3}}.
\end{equation}
{ where $\Gamma$ is a solution of the $\ov\pa$-problem \ref{eq:NLShdbar}}  so that we obtain  the $\ov\pa$-problem
\begin{equation}
  \label{eq:new_d_bar}
  \left\{
    \begin{split}
      \pa_{\bar{z}}\Psi(z)&=\Psi(z)M_{0}(z)\\
      \Psi(z)&=\left( \1 + \mathcal{O}\left( z^{-1}\right) \right){\rm e}^{-i(zx  +z^2 t)\sigma_{3}}  \quad \text{as } z \to \infty\,.
    \end{split} \right.
\end{equation}
We denote the terms of the expansion of $\Psi$ near $z=\infty$ as follows:
\be
\label{psiexp}
\Psi(z;x,t) = \le(\1 +\sum_{\ell=1}^\infty \frac {\Gamma_\ell(x,t)}{z^\ell} \ri){\rm e}^{-i(zx  +z^2 t)\sigma_{3}}\,.
\ee
The first observation is that $\Psi$ satisfies the Schwartz-like symmetry 
\be
\label{SchwartzPsi}
{\Psi(z;x,t)}^\dagger \Psi(\ov z;x,t) \equiv \1,
\ee
which follows from the uniqueness of the solution after observing that the matrix $\Phi(z;x,t):= {\Psi(z;x,t)}^\dagger $ solves the same $\ov\pa$-problem, thanks to the property $M(z;x,t) =- M(\ov z;x,t)^\dagger$. Given that $\det \Psi\equiv 1$ we can rewrite the symmetry as 
\be
\Psi(z;x,t) = \sigma_2 \Psi(\ov z;x,t)^\dagger \sigma_2. \label{ShcwartzPsi2}
\ee
This translates to the following symmetry for the matrices $\Gamma_\ell(x,t)$:
\be
\label{SchwartzGamma}
\Gamma_\ell(x,t) = \sigma_2 \ov{\Gamma_\ell(x,t)}\sigma_2.
\ee
Since the operators $\partial_{x}$ and $\pa_{\bar{z}}$ commute, we can see that $\partial_{x}\Psi$ satisfies the problem \eqref{eq:new_d_bar}
\begin{equation}
  \pa_{\bar{z}}\pa_x \Psi= \pa_x\Psi M_{0}(z).
\end{equation}
It now follows that the matrix   $U(z;x,t):= \pa_x\Psi(\Psi^{-1})$ is an entire function in $z$. Indeed
\begin{equation*}
  \begin{split}
    \partial_{\bar{z}}\left( \pa_x \Psi\Psi^{-1}\right)&= (\partial_{\bar{z}}\pa_x\Psi)\Psi^{-1} + \pa_x\Psi(\pa_{\bar{z}}\Psi^{-1})\\
    &=(\pa_x \Psi) M_{0}\Psi^{-1} -(\pa_x \Psi)\Psi^{-1}\partial_{\bar{z}}\Psi (\Psi^{-1})\\
    &=(\pa_x \Psi) M_{0}\Psi^{-1} - (\pa_x \Psi) M_{0}\Psi^{-1}=0.
  \end{split}
\end{equation*}
Thus we obtain the   following equation:
\begin{equation}
  \label{eq:Lax_1}
  \pa_x \Psi(z;x,t)=U(z;x,t)\Psi(z;x,t).
\end{equation}
In order to  determine the $z$--dependence of $U(z;x,t)$ we  consider the asymptotic of $\Psi(z;x,t)$ for $z \to \infty$  by differentiation of the asymptotic behaviour specified in \eqref{eq:new_d_bar}  
\begin{equation*}
  \begin{split}
\pa_x \Psi \sim -\left( \1 +\frac{\Gamma_{1}(x,t)}{z} + \mathcal{O}\left(z^{-2}\right) \right)iz\sigma_{3}{\rm e}^{-i(zx  +z^2 t) \sigma_{3}} +\left(\frac{\partial_{x}\Gamma_{1}(x,t)}{z}  + \mathcal{O}\left( z^{-2}\right)\right){\rm e}^{-i(zx  +z^2 t)\sigma_{3}}   .
  \end{split}
\end{equation*}
Upon substitution in~\eqref{eq:Lax_1} we get
\begin{equation}
  \label{eq:U}
  \begin{split}
    U(z;x,t) &= \pa_x \Psi (\Psi^{-1} )\sim\\
    &\sim -iz\left( \mathbb{I} + \frac{\Gamma_{1}(x,t)}{z} \right)\sigma_{3}\left( \1 -\frac{\Gamma_{1}(x,t)}{z} \right) + \mathcal{O}\left(z^{-1}\right)\\
    &= -iz\sigma_{3} -i\left[\Gamma_{1}(x,t),\sigma_{3}\right] + \mathcal{O}\left(z^{-1}\right).
  \end{split}
\end{equation}
and since we know that $U(z;x,t)$ is entire, we conclude that $U$ is the polynomial in $z$ of first degree obtained by dropping the $\mathcal O(z^{-1})$ in \eqref{eq:U}. Due to the symmetry \eqref{SchwartzGamma} the matrix $\Gamma_{1}(x,t)$ has the  form
\begin{equation}
  \label{eq:Gamma_1}
  \Gamma_{1}(x,t)=
  \begin{bmatrix}
    a(x,t) & b(x,t)\\
    -\bar{b}(x,t) & \bar{a}(x,t)
  \end{bmatrix}, 
\end{equation}
from which we find  
\begin{equation*}
  \left[\Gamma_{1}(x,t),\sigma_{3}\right]=
  \begin{bmatrix}
0 & -2b(x,t)\\
    -2\bar{b}(x,t) & 0
  \end{bmatrix}\,.
\end{equation*}
We thus conclude that the matrix $U(z;x,t)$ has the form
\begin{equation}
  \label{eq:U_1}
  U(z;x,t)=
  \begin{bmatrix}
    -iz & 2ib(x,t)\\
    2i\bar{b}(x,t) & iz
  \end{bmatrix}.
\end{equation}
The same arguments can be applied  for the parameter $t$. In that case, $\Psi(z;x,t)$ satisfy the ODE
\begin{equation}
  \pa_t \Psi(z;x,t)=V(z;x,t)\Psi(z;x,t)
\end{equation}
where $V(z;x,t)$ is an entire function in $z$.
Following the same idea as before, we expand $\pa_t \Psi(z;x,t)$ for $z\to \infty$
\begin{equation}
  \pa_t \Psi \sim- iz^2\left(\mathbb{I} + \frac{\Gamma_{1}(x,t)}{z} + \frac{\Gamma_{2}(x,t)}{z^{2}} + \mathcal{O}\left(z^{-3}\right)\right)\sigma_{3}{\rm e}^{-i(zx  +z^2 t)\sigma_{3}} + \mathcal{O}\left(z^{-1}\right){\rm e}^{-i(zx  +z^2 t)\sigma_{3}}
\end{equation}
and we have
\begin{equation}
  \begin{split}
    \pa_t \Psi(\Psi^{-1})&=V(z;x,t) \sim -\left(\mathbb{I} +\frac{\Gamma_{1}(x,t)}{z} + \frac{\Gamma_{2}(x,t)}{z^{2}} +\mathcal{O}\left(z^{-3}\right)  \right)\times\\
    & \times iz^{2}\sigma_{3}\left( \mathbb{I} -\frac{\Gamma_{1}(x,t)}{z} -\frac{\Gamma_{2}(x,t)}{z^{2}} +\frac{(\Gamma_{1}(x,t))^{2}}{z^{2}} + \mathcal{O}\left(z^{-3}\right)  \right)\\
    &= -iz^{2}\sigma_{3} - iz\left[\Gamma_{1}(x,t),\sigma_{3}\right] - i\left[\Gamma_{2}(x,t),\sigma_{3}\right] +i\left[\Gamma_{1}(x,t),\sigma_{3}\right]\Gamma_{1} (x,t)+ \mathcal{O}\left(z^{-1}\right).
  \end{split}
\end{equation}
We similarly conclude that $V(z;x,t)$ is the polynomial part of the above expression, a quadratic polynomial in $z$.
To complete the calculation we need to relate  the matrix $\Gamma_{2}(x,t)$ to the $\pa_x$ derivative of $\Gamma_1$ by taking  the expansion of both sides of the Lax equation \eqref{eq:Lax_1} as $z\to\infty$, and using the explicit expression of $U$ given in \eqref{eq:U}. 
The term $\mathcal{O}(z^{-1})$ in~\eqref{eq:Lax_1} provides the equation:
\begin{equation}
\label{eq:z_1}
  \pa_{x} \Gamma_{1}(x,t)= i [\Gamma_{2}(x,t),\sigma_{3}] -i[\Gamma_{1}(x,t),\sigma_3]\Gamma_{1}(x,t) .
\end{equation}
%
%
The $(1,1)$ entry of~\eqref{eq:z_1} yields the relation
\begin{equation}
  \label{eq:rec_comp1}
  \pa_{x} a(x,t)=-2i|b(x,t)|^{2}
\end{equation}
while the off diagonal give
\begin{equation}
  (\left[\Gamma_{2}(x,t),\sigma_{3}\right])_{12}=(\overline{\left[\Gamma_{2}(x,t),\sigma_{3}\right]})_{21}= -2{b}\ov {a} - i \pa_xb.
\end{equation}
In conclusion,  the matrix $V(z;x,t)$ is
\begin{equation}
  \begin{split}
    V(z;x,t)&= -iz^{2}\sigma_{3} - iz\left[\Gamma_{1}(x,t),\sigma_{3}\right] - i\left(\left[\Gamma_{2}(x,t),\sigma_{3}\right] - \left[\Gamma_{1}(x,t),\sigma_{3}\right]\Gamma_1(x,t)\right)\\
    &= -iz^{2}\sigma_{3} - iz\left[\Gamma_{1}(x,t),\sigma_{3}\right] -\partial_{x}\Gamma_{1}(x,t)\\
    &=
    \begin{bmatrix}
      -iz^{2}+2i|b|^{2} & 2zb -\pa_xb\\
      2z\bar{b} +\pa_x\bar{b} & iz^{2} -2i|b|^{2}
    \end{bmatrix}.
  \end{split}
\end{equation}
Summarizing, the matrix $\Psi(z;x,t)$  solves the $\bar{\partial}$-Problem~\ref{eq:new_d_bar} as well as the two linear PDEs 
\begin{equation}
  \label{eq:Lax_pair}
  \begin{split}
    \pa_x \Psi &= U(z;x,t) \Psi = 
    \begin{bmatrix}
      -iz & \psi\\
      -\bar{\psi} & iz
    \end{bmatrix}\Psi\\
    \pa_t \Psi &= V(z;x,t)\Psi=
    \begin{bmatrix}
      -iz^{2} +\frac{i}{2}|\psi|^{2} & z\psi  +\frac{i}{2}\pa_x\psi\\
      -z\bar{\psi} +\frac{i}{2}\pa_x\bar{\psi} & iz^{2} - \frac{i}{2}|\psi|^{2}
    \end{bmatrix}\Psi
  \end{split}
\end{equation}
where we have set $\psi(x,t):=2ib(x,t)$. We   can see that the matrices $U(z;x,t)$ and $V(z;x,t)$ are in the form of the Lax pair of the NLS~\eqref{eq:NLS},  namely, the zero curvature equations  \cite{ZS}
\be
\pa_x\pa_t\Psi \equiv \pa_t\pa_x\Psi \ \ \Leftrightarrow\ \ \ 
\pa_t U- \pa_x V + [U,V]\equiv 0
\ee
and the latter is equivalent to the NLS equation~\eqref{eq:NLS}.
\section{Proof of Lemma~\ref{lemm-2}}
\label{app-II}
In this section we give the proof of Lemma~\ref{lemm-2}.
Since the computations of  $\omega(\pmb{t} \pm [\zeta^{-1}])$ are the same, we give the proof only for $\omega(\pmb{t} - [\zeta^{-1}])$.

From~\eqref{eq:Malg_form},~\eqref{eq:M_s} and~\eqref{eq:Gamma-1}, we get
  \bea
       \omega(&\pmb{t} -[\zeta^{-1}])=-\iint_{\mathscr{D}}\Tr\le(\Gamma^{-1}\le(z,\pmb{t} -[\zeta^{-1}]\ri)\pa_z\Gamma\le(z,\pmb{t} -[\zeta^{-1}]\ri)\delta_{[\zeta]} M\le(z,\pmb{t} -[\zeta^{-1}]\ri)\ri)\frac{\d \ov z\wedge \d {z}}{2 \pi i} \nn \\ 
      =&-\iint_{\mathscr{D}}\Tr\le(D(z)\Gamma^{-1}(z)C^{-1}(z)\pa_z
     \le(C(z)\Gamma(z)D^{-1}(z)\ri)\delta_{[\zeta]}\!\!\le(D(z)M(z,\pmb{t})D^{-1}(z)\ri)\ri)\frac{ \d \ov{z} \wedge \d z}{2\pi i} \nonumber\\
     =&-\iint_{\mathscr{D}}\Tr\le(D(z)\Gamma^{-1}(z)\pa_z\Gamma(z)D^{-1}(z)\delta_{[\zeta]}\le(D(z)M(z,\pmb{t})D^{-1}(z)\ri)\ri)\frac{ \d \ov{z} \wedge \d z}{2\pi i}  + \label{eq:A}\\
     &-\iint_{\mathscr{D}}\Tr\le(D(z)\Gamma^{-1}(z)C^{-1}(z)\pa_z C(z)\Gamma(z)D^{-1}(z)\delta_{[\zeta]}\le(D(z)M(z,\pmb{t})D^{-1}(z)\ri)\ri)\frac{ \d \ov{z} \wedge \d z}{2\pi i} + \label{eq:B}\\
      &-\iint_{\mathscr{D}}\Tr\le(D(z)\pa_zD^{-1}(z)\delta_{[\zeta]}\le(D(z)M(z,\pmb{t})D^{-1}(z)\ri)\ri)\frac{ \d \ov{z} \wedge \d z}{2\pi i} \label{eq:C}
  \eea
 
  We now consider the three parts \eqref{eq:A}, \eqref{eq:B}, \eqref{eq:C}, separately.
\paragraph{Computation of \eqref{eq:A}.}
We find:
  \begin{equation*}
    \begin{split}
      \eqref{eq:A}&= -\iint_{\mathscr{D}}\Tr\le(\Gamma^{-1}(z)\pa_z\Gamma(z)\delta M(z,\pmb{t})\ri)\frac{ \d \ov{z} \wedge \d z}{2\pi i} + \\
      &-\iint_{\mathscr{D}}\Tr\le(\Gamma^{-1}(z)\pa_z\Gamma(z)
      \bigg[D^{-1}(z)\delta_{\zeta} D(z), M(z,\pmb{t})\bigg]
      \ri)\frac{ \d \ov{z} \wedge \d z}{2\pi i}%
\\
      &=\omega(\pmb{t}) +
     \iint_{\mathscr{D}}\Tr\le(D^{-1}(z)\delta_{\zeta} D(z)
      \bigg[ \Gamma^{-1}(z)\pa_z\Gamma(z),M(z,\pmb{t})\bigg]
      \ri)\frac{ \d \ov{z} \wedge \d z}{2\pi i}.
    \end{split}
  \end{equation*}
  Since $\zeta \notin \mathscr{D}$, the matrix $D^{-1}(z)$ in \eqref{eq:G-shift} is analytic in $\mathscr{D}$ and  using the $\ov\pa$-problem for $\Gamma$ we can rewrite the two integrals as
  \begin{equation}
    \label{eq:A_mid}
    \begin{split}
      \eqref{eq:A} &= \omega(\pmb{t}) 
      + \iint_{\mathscr{D}}\pa_{\bar{z}}\Tr\bigg(\Gamma^{-1}(z)\pa_z\Gamma(z) D^{-1}(z)\delta_{\zeta} D(z)\bigg)\frac{ \d \ov{z} \wedge \d z}{2\pi i} +\\
      &- \iint_{\mathscr{D}}\Tr\le(\pa_z M(z,\pmb{t})D^{-1}(z)\delta_{\zeta} D(z)\ri)\frac{ \d \ov{z} \wedge \d z}{2\pi i}.
    \end{split}
  \end{equation}
  We now observe that the last integral is independent of $\pmb{t}$, due to the fact that $D(z)$ is diagonal. Moreover, using
  \begin{equation*}
    D^{-1}(z)\delta_{\zeta} D(z)= -\frac{z}{\zeta(z-\zeta)} E_{11}\d \zeta,
  \end{equation*}
   where $E_{11}= \begin{bmatrix} 1 & 0\\ 0 & 0\end{bmatrix}$, we find 
  \begin{equation}
    \label{eq:LI}
 - \iint_{\mathscr{D}} \Tr\le(\pa_z M(z,\pmb{t}) D^{-1}(z)\pa_{\zeta} D(z)\ri)\frac{ \d \ov{z} \wedge \d z}{2\pi i}= \iint_{\mathscr{D}}\frac{z}{\zeta(z-\zeta)}(\pa_z M_{0}(z))_{11} \frac{ \d \ov{z} \wedge \d z}{2\pi i}.
  \end{equation}
 The RHS of~\eqref{eq:LI} equals $\pa_{\zeta} \gamma(\zeta)$.
 Now, the integrand of the remaining integral in~\eqref{eq:A_mid} does not have a pole in $\mathscr{D}$ and  we can use Stokes' Theorem 
  \begin{equation*}
    \oint_{\pa \mathscr{D}}\Tr
    \le(\Gamma^{-1}(z)\pa_z\Gamma(z)D^{-1}(z)\pa_{\zeta} D(z)\ri)
    \frac{\d z}{2\pi i}=\oint_{-\pa \mathscr{D}}\frac{z}{\zeta(z-\zeta)}(\Gamma^{-1}(z)\pa_z\Gamma(z))_{11}\frac{\d z}{2 \pi i}
  \end{equation*}
  where $-\pa \mathscr{D}$ is the border of $\mathscr{D}$ oriented clockwise.
   Since $\Gamma(z)$ is analytic outside $\mathscr{D}$, we can apply Cauchy's residue Theorem  and pick up the residues at $z=\zeta$ (there is no residue at $z=\infty$ because the integrand is $\mathcal O(z^{-2})$):
  \begin{equation*}
 \oint_{-\pa \mathscr{D}}\frac{z}{\zeta(z-\zeta)}\big(\Gamma_{22}(z)\pa_z\Gamma_{11}(z) -\pa_z\Gamma_{21}(z)\Gamma_{12}(z)\big)\frac{\d z}{2 \pi i}= \Gamma_{22}(\zeta)\pa_\zeta\Gamma_{11}(\zeta) -\pa_\zeta\Gamma_{21}(\zeta)\Gamma_{12}(\zeta)   
\end{equation*}
so that
\begin{equation}
  \label{eq:Af}
  \eqref{eq:A}= \omega(\pmb{t}) 
  +\left( \Gamma_{22}(\zeta)\pa_{\zeta}\Gamma_{11}(\zeta) - \pa_{\zeta}\Gamma_{21}(\zeta)\Gamma_{12}(\zeta)\right) \d \zeta + \delta_{\zeta} \gamma(\zeta).
\end{equation}
\paragraph{Computation of \eqref{eq:B}.}
Let us consider \eqref{eq:B}: 
\bea
\nn   \eqref{eq:B} &=-\iint_{\mathscr{D}}
\Tr
\le(\Gamma^{-1}(z)C^{-1}(z)\pa_z C(z)\Gamma(z)\delta M(z,\pmb{t})\ri)\frac{ \d \ov{z} \wedge \d z}{2\pi i}+\\
 \nn
    &-\iint_{\mathscr{D}}\Tr
    \le(M(z,\pmb{t})\Gamma^{-1}(z)C^{-1}(z)\pa_z C(z)\Gamma(z)D^{-1}(z) \delta_{\zeta} D(z)\ri)\frac{ \d \ov{z} \wedge \d z}{2\pi i} +\\
 \nn
    &+\iint_{\mathscr{D}}\Tr
    \le(\Gamma^{-1}(z)C^{-1}(z)\pa_z C(z)\Gamma(z)M(z,\pmb{t})D^{-1}(z)\delta_{\zeta} D(z)\ri)\frac{ \d \ov{z} \wedge \d z}{2\pi i}\\
 \nn
    &=-\iint_{\mathscr{D}}\Tr
    \le(\Gamma^{-1}(z)C^{-1}(z)\pa_z C(z)\Gamma(z)\delta M(z,\pmb{t})\ri)
    \frac{ \d \ov{z} \wedge \d z}{2\pi i}+\\
 \nn
    &+\iint_{\mathscr{D}}\Tr\le(\pa_{\bar{z}}\Gamma^{-1}(z)C^{-1}(z)\pa_z C(z)\Gamma(z)D^{-1}(z) \delta_{\zeta} D(z)\ri)
    \frac{ \d \ov{z} \wedge \d z}{2\pi i}+\\
    &+\iint_{\mathscr{D}}\Tr
    \le(\Gamma^{-1}(z)C^{-1}(z)\pa_z C(z)\ov{\pa}\Gamma(z)D^{-1}(z)\delta_{\zeta} D(z)\ri)\frac{ \d \ov{z} \wedge \d z}{2\pi i}. 
    \label{B7}
\eea
Since the only singularity is at $z=\zeta$, which is outside the domain $\mathscr{D}$, we can apply Stokes' Theorem to the integration and we get
\bea\nn
   \eqref{eq:B} &=-\iint_{\mathscr{D}}\Tr
   \le(C^{-1}(z)\pa_z C(z)\Gamma(z)\delta M(z,\pmb{t})\Gamma^{-1}(z)\ri)\frac{ \d \ov{z} \wedge \d z}{2\pi i} +\\
    &+\oint_{\pa \mathscr{D}}\Tr\le(
    \Gamma^{-1}(z)C^{-1}(z)\pa_z C(z)\Gamma(z)D^{-1}(z)\delta_{\zeta} D(z)\ri)\frac{\d z}{2\pi i}.
\label{B8}
\eea
Now observe that
\begin{equation}
  \label{eq:I_int_B}
  \Gamma(z,\pmb{t}) \delta M(z,\pmb{t}) \Gamma^{-1}(z,\pmb{t})=\pa_{\bar{z}}[(\delta \Gamma(z,\pmb{t})) \Gamma^{-1}(z,\pmb{t})]
\end{equation}
Using~\eqref{eq:I_int_B} in the first integral of \eqref{B8}, we can rewrite it as a contour integral
\begin{equation*}
  \begin{split}
    &-\iint_{\mathscr{D}}\Tr\le(
    \Gamma^{-1}(z)C^{-1}(z)\pa_z C(z)\Gamma(z)\delta M(z,\pmb{t})\ri)
    \frac{ \d \ov{z} \wedge \d z}{2\pi i} \\
    &=-\iint_{\mathscr{D}}\pa_{\bar{z}}\Tr
    \le(C^{-1}(z)\pa_z C(z)\delta \Gamma(z) \Gamma^{-1}(z)\ri)\frac{ \d \ov{z} \wedge \d z}{2\pi i}\\
    &= \oint_{-\pa \mathscr{D}}\Tr\le(C^{-1}(z)\pa_z C(z)\delta \Gamma(z) \Gamma^{-1}(z)\ri)\frac{\d z}{2\pi i}.
    \end{split}
  \end{equation*}
From the explicit expression of $C$ in \eqref{eq:R-mat} we obtain
  \begin{equation*}
    C^{-1}(z)= \frac{1}{\det C(z)} {\rm adj}(C(z))= \frac{1}{\left(1 - \frac{z}{\zeta}\right)}
    \begin{bmatrix}
      1 & -\frac{\pa_z\Gamma_{12}(\infty)}{\zeta}\\
      \frac{\Gamma_{21}(\zeta)}{\Gamma_{11}(\zeta)} & \left(1 - \frac{z}{\zeta}\right) - \frac{\pa_z \Gamma_{12}(\infty) \Gamma_{12}(\zeta)}{\zeta\Gamma_{11}(\zeta)}
    \end{bmatrix}
  \end{equation*}
  \begin{equation*}
    \pa_z C(z) = -\frac{1}{\zeta} E_{11}
  \end{equation*}
  and
  \begin{equation*}
    \begin{split}
      &\Tr\le(C^{-1}(z)\pa_z C(z)\delta\Gamma(z) \Gamma^{-1}(z)\ri)=\frac{1}{(z-\zeta)}\left((\delta \Gamma(z)\Gamma^{-1}(z))_{11} + \frac{\Gamma_{21}(\zeta)}{\Gamma_{11}(\zeta)}(\delta \Gamma(z)\Gamma^{-1}(z))_{12}\right)\\
      &=\frac{\delta\Gamma_{11}(z)\Gamma_{22}(z) - \delta \Gamma_{12}(z)\Gamma_{21}(z)}{(z - \zeta)} + \frac{\Gamma_{21}(\zeta)}{\Gamma_{11}(\zeta)}\left(\frac{\delta \Gamma_{12}(z)\Gamma_{11}(z) - \delta \Gamma_{11}(z) \Gamma_{12}(z)}{(z - \zeta)}\right).
    \end{split}
  \end{equation*}
  We thus conclude that  the first integral in~\eqref{B8} is given by 
\bea
      \oint_{-\pa \mathscr{D}}\Tr
      \big(C^{-1}(z)\pa_z C(z)&\delta \Gamma(z) \Gamma^{-1}(z)\big)\frac{\d z}{2\pi i} = \delta \Gamma_{11}(\zeta)\Gamma_{22}(\zeta) - \frac{\delta \Gamma_{11}(\zeta)}{\Gamma_{11}(\zeta)}\Gamma_{12}(\zeta)\Gamma_{21}(\zeta)\nn
      \\
      &= \frac{\delta \Gamma_{11}(\zeta)}{\Gamma_{11}(\zeta)}= \delta \ln \Gamma_{11}(\zeta).\label{B9}
\eea
  To compute the second integral in \eqref{B8} we expand the trace and obtain 
  \begin{equation*}
    \Tr\big(\Gamma^{-1}(z)C^{-1}(z)\pa_z C(z)\Gamma(z)D^{-1}(z)\pa_{\zeta} D(z)\big)= -\frac{z}{\zeta(z-\zeta)^{2}}\Gamma_{11}(z)\left(\Gamma_{22}(z) - \frac{\Gamma_{12}(z)\Gamma_{21}(\zeta)}{\Gamma_{11}(\zeta)}\right).
  \end{equation*}
  So we are left with a contour integral with a double pole at $z=\zeta$ and a simple pole at $z=\infty$. Using the explicit expression \eqref{eq:R-mat} for the matrix $C$ we obtain: 
  \bea\nn
        \oint_{\pa \mathscr{D}}&\Tr
        \le(\Gamma^{-1}(z)C^{-1}(z)\pa_z C(z)\Gamma(z)D^{-1}(z)\pa_\zeta D(z)\ri)\frac{\d z}{2\pi i}\\
      \nn
      &=\oint_{-\pa \mathscr{D}}\frac{z}{\zeta(z-\zeta)^{2}}\Gamma_{11}(z)\left(\Gamma_{22}(z) - \frac{\Gamma_{12}(z)\Gamma_{21}(\zeta)}{\Gamma_{11}(\zeta)}\right)\frac{\d z}{2\pi i}\nn
\\
\nn
&=-\frac 1 \zeta 
  + \frac {\det \Gamma(\zeta)}{\zeta} +\pa_\zeta(\Gamma_{11}(\zeta)\Gamma_{22}(\zeta))  
  -\frac{  \pa_\zeta\big(\Gamma_{11}(\zeta)\Gamma_{12}(\zeta)\big)\Gamma_{21}(\zeta)}{\Gamma_{11}(\zeta)}
  \\
  &=\pa_\zeta\ln \Gamma_{11}(\zeta)
  +  \Gamma_{11}(\zeta) \pa_\zeta\Gamma_{22}(\zeta)  - \pa_\zeta\Gamma_{12}(\zeta) \Gamma_{21}(\zeta).
  \label{B10}
      \eea
Combining \eqref{B9} with \eqref{B10}  we have 
  \begin{equation}
    \label{eq:Bf}
    \eqref{eq:B}= \delta_{[\zeta]}(\ln(\Gamma_{11}(\zeta))) +\left( \Gamma_{11}(\zeta)\pa_{\zeta}\Gamma_{22}(\zeta) - \Gamma_{21}(\zeta)\pa_{\zeta}\Gamma_{12}(\zeta) \right)\d \zeta.
  \end{equation}
  \paragraph{Computation of \eqref{eq:C}.}
 This term turns out to vanish; indeed
  \begin{equation*}
    \begin{split}
      \eqref{eq:C}&= \iint_{\mathscr{D}}\Tr
      \le(D^{-1}(z)\pa_z D(z)\delta M(z,\pmb{t})\ri)
      \frac{ \d \ov{z} \wedge \d z}{2\pi i}+\\
      &\quad-\iint_{\mathscr{D}}\Tr
      \le(D^{-1}(z)\pa_z D(z)\le[M(z,\pmb{t}), D^{-1}(z)\delta_{\zeta} D(z)\ri]\ri)\frac{ \d \ov{z} \wedge \d z}{2\pi i}\\
      &= -\iint_{\mathscr{D}}\Tr
      \le(\delta\xi(z,t)D^{-1}(z)\pa_z D(z) \le[M(z,\pmb{t}),\sigma_{3}\ri]\ri)\frac{ \d \ov{z} \wedge \d z}{2\pi i} +\\
      &\quad-\iint_{\mathscr{D}}\Tr
      \le(D^{-1}(z)\pa_z D(z)\le[M(z,\pmb{t}),D^{-1}(z)\delta_{\zeta} D(z)\ri]\ri)\frac{ \d \ov{z} \wedge \d z}{2\pi i}
      \end{split}
    \end{equation*}
 and the integrand vanishes identically because of the cyclicity of the trace and the fact that $D$ is a diagonal matrix.
    In conclusion,  adding the equations~\eqref{eq:Af} and~\eqref{eq:Bf}, we obtain
    \begin{equation}
      \label{eq:res-p1}
      \omega(\pmb{t}-[\zeta^{-1}])=\omega(\pmb{t}) 
      + \delta_{[\zeta]} \ln(\Gamma_{11}(\zeta)) + \delta_{[\zeta]} \gamma(\zeta) .
    \end{equation}
    Substituting $C(z)$ and $D(z)$ with $\tilde{C}(z)$ and $\tilde{D}(z)$ respectively and using the nonsingular condition for $K$~\eqref{eq:K_ker2}, we find $\omega(\pmb{t} + [\zeta^{-1}])$ with similar calculations and we get the following result
    \begin{equation}
      \label{eq:res-p2}
      \omega(\pmb{t}+[\zeta^{-1}])= \omega(\pmb{t}) + \delta_{[\zeta]}\ln(\Gamma^{-1}_{11}(\zeta)) - \delta_{[\zeta]}\gamma(\zeta).
    \end{equation}
    and this proves the Lemma~\ref{lemm-2}.\QED

\bibliographystyle{siam}

 \end{document}